\documentclass{emulateapj}
\pdfoutput=1
\usepackage{epsfig,natbib}
\usepackage{graphicx}
\usepackage{enumerate}
\usepackage{amsmath}
\usepackage{multirow}
\usepackage{bigdelim}
\usepackage{longtable}
\usepackage{threeparttable}
\usepackage{subfigure}
\usepackage{textcomp}
\usepackage{gensymb}
\usepackage{color}
\usepackage{placeins}

\newcommand{\msun}{M$_{\odot}~$}

\newcommand{\Rmnum}[1]{\expandafter\@slowromancap\romannumeral #1@}

\newcommand{\vect}[1]{\boldsymbol{#1}}
\newcommand{\murel}{\mu_{\mathrm{rel}}}
\newcommand{\murelvec}{\vect{\mu}_{\boldsymbol{\mathrm{rel}}}}
\newcommand{\thetaEhat}{\boldsymbol{\hat{\theta}}_{\boldsymbol{E}}}

\newcommand{\thetaevec}{\vect{\theta}_{\boldsymbol{E}}}
\newcommand{\uvec}{\vect{u}}
\newcommand{\muSvec}{\vect{\mu}_{\boldsymbol{s}}}
\newcommand{\uveco}{\vect{u}_{\boldsymbol{0}}}
\newcommand{\deltavec}{\vect{\delta}_{\boldsymbol{c}}}
\newcommand{\OBtwotwo}{OGLE-2011-BLG-0022}
\newcommand{\OBtwofive}{OGLE-2011-BLG-0125}
\newcommand{\OBsixnine}{OGLE-2012-BLG-0169}
\newcommand{\tE}{t_E}
\newcommand{\thetaE}{\theta_E}
\newcommand{\uo}{u_0}
\newcommand{\uoplus}{u_0^+}
\newcommand{\uominus}{u_0^-}

\newcommand{\thetastar}{\theta_{\star}}

\newcommand{\piEvec}{\vect{\pi}_E}
\newcommand{\piE}{\pi_E}

\newcommand{\tnaught}{t_0}
\newcommand{\fsrc}{f_{s}}
\newcommand{\fb}{f_{b}}
\newcommand{\piEN}{\pi_{E, N}}
\newcommand{\piEE}{\pi_{E, E}}

\newcommand{\subsun}{_{\odot}}
\newcommand{\x}{E}
\newcommand{\y}{N}
\newcommand{\sigmaX}{\sigma_{\x, i}}
\newcommand{\sigmaY}{\sigma_{\y, i}}
\newcommand{\xpos}{x_{\x}}
\newcommand{\ypos}{x_{\y}}
\newcommand{\xobs}{x_\x^\mathrm{obs}}
\newcommand{\yobs}{x_\y^\mathrm{obs}}
\newcommand{\tobs}{t^\mathrm{obs}}
\newcommand{\xobsi}{x_{\x, i}^{\mathrm{obs}}}
\newcommand{\yobsi}{x_{\y, i}^{\mathrm{obs}}}

\newcommand{\xmodi}{x_{\x, i}^{\mathrm{mod}}}
\newcommand{\ymodi}{x_{\y, i}^{\mathrm{mod}}}

\newcommand{\Xsovec}{\vect{X}_{\boldsymbol{{0}}}}
\newcommand{\Xsvec}{\vect{X}}

\shortauthors{Lu et al.}
\begin{document}
\pagenumbering{arabic}

\title{A search for stellar-mass black holes via astrometric microlensing \\
}
\author{
J.~R.~Lu\altaffilmark{1},
E. Sinukoff\altaffilmark{1},
E.~O.~Ofek\altaffilmark{2},
A. Udalski\altaffilmark{3},
S. Kozlowski\altaffilmark{3}
}

\altaffiltext{1}{Institute for Astronomy,
University of Hawai`i at M\={a}noa, Honolulu, HI 96822}
\altaffiltext{2}{Department of particle physics and astrophysics
Weizmann Institute of Science, Rehovot, Israel 76100}
\altaffiltext{3}{Warsaw University Observatory, Al. Ujazdowskie 4, 00-478 Warszawa, Poland}

\begin{abstract}
While dozens of stellar mass black holes have been discovered in
binary systems, isolated black holes have eluded detection.  
Their presence can be inferred when they lens light from a background
star.  We attempt to detect the astrometric lensing signatures of
three photometrically identified microlensing events, \OBtwotwo,
\OBtwofive, and \OBsixnine~(OB110022, OB110125, and OB120169), located
toward the Galactic Bulge.  These events were selected because of
their long durations, which statistically favors more massive lenses.
Astrometric measurements were made over 1--2 years
using laser-guided adaptive optics observations from the
W. M. Keck Observatory.
Lens model parameters were first constrained by the
photometric light curves.
The OB120169 light curve is well-fit by a single-lens model, while
both OB110022 and OB110125 light curves favor binary-lens models.
Using the photometric fits as prior information,
no significant astrometric lensing signal was detected and all
targets were consistent with linear motion.
The significant lack of astrometric signal constrains the
%
%
lens mass of OB110022 to 0.05--1.79 M$\subsun$ in a 99.7\% confidence
interval, which disfavors a black hole lens.
Fits to OB110125 yielded a reduced Einstein crossing time and 
insufficient observations during the peak, so no
mass limits were obtained. 
Two degenerate solutions exist for OB120169, which have a lens mass
%
%
between 0.2--38.8 M$\subsun$ and 0.4--39.8 M$\subsun$ for a 99.7\%
confidence interval.  Follow-up observations of OB120169
will further constrain the lens mass.
Based on our experience, we use simulations to design
optimal astrometric observing strategies and show that, with more
typical observing conditions, detection of black holes is feasible.

\end{abstract}

\keywords{astrometry --- gravitational lensing: micro --- stars: black
  holes --- instrumentation: adaptive optics}

\section{Introduction}
\label{sec:intro}

Core-collapse supernova events, which mark the deaths of high-mass
($\gtrsim$ 8 M$\subsun$) stars, are predicted to leave remnant black
holes of order several to tens of M$\subsun$. It is estimated that
10$^{8}$--10$^{9}$ of these ``stellar mass black holes" occupy the
Milky Way Galaxy \citep{Agol02,Gould:2000}.
Detecting isolated black holes (BHs) and measuring their masses
constrains the number density and mass function of
BHs within our Galaxy. These factors have important implications
for how BHs form, supernova physics, and the equation of state of
nuclear matter. For example, the BH mass function can be compared to
the stellar initial mass function to define
the initial-final mass relation including which stars produce BHs rather than neutron stars.
Such measurements can help test different supernova explosion
mechanisms, which predict different initial-final mass relations, and
constrain the fraction of ``failed supernova explosions'' that lead to
BH formation
\citep[e.g.][]{Gould2002,Kochanek:2008,KushnirKatz:2015,Kushnir:2015,Pejcha:2015}.
Additionally, the BH occurrence rate is a key input into predictions
for future BH detection missions like the Laser Interferometer Space
Antenna \citep[LISA,][]{Prince07} and the the Laser Interferometer
Gravitational-Wave Observatory \citep[LIGO,][]{Abbott09}.

To date, a few dozen BHs have been detected, but discoveries have been
limited to BHs in binary systems.  All of these are actively accreting
from a binary companion and emitting strongly at radio or X-ray
wavelengths \citep[see e.g.][for a review]{Reynolds13,
  Casares14}. Isolated BHs, which could comprise the majority of the
BH population, remain elusive, with no confirmed detections to date.
These objects can only accrete from the surrounding interstellar
medium, producing minimal emission presumably in soft X-rays.

While isolated BHs do not produce detectable emission of their own,
their gravity can noticeably bend and focus (i.e.~lens) light from any
background source in close proximity on the sky, allowing their
presence to be inferred.   During these chance alignments, the
relative proper motion between the source and lens,
$\boldsymbol{\vec{\mu}_{\mathrm{rel}}}$ produces a transient event
that is observable both photometrically and astrometrically with the
following signatures: (1) The background source increases in apparent
brightness, and (2) The position of the source shifts astrometrically
and splits into multiple images
\citep{Miyamoto:1995,Hog:1995,Walker:1995}.
For stellar-mass lensing events in the Galaxy, these multiple images
are unresolved with current telescopes and are deemed ``microlensing"
events \citep{Paczynski86}.  Photometric microlenses are frequently
detected by large transient surveys such as the Optical Gravitational
Lensing Experiment \citep[OGLE,][]{Udalski92} and the Microlensing
Observations in
Astrophysics survey \citep[MOA,][]{Bond01}.  However, astrometric
shifts from a microlensing event have never been detected. Depending
on the mass and relative distances to the source and lens, the
astrometric shift of a BH lens is $\sim$1 mas and the event duration
can be months to years.  Although the successful detection of such
astrometric signatures is challenging, it would have significant
payoff; astrometric information can be combined with photometric
measurements to precisely constrain lens masses. 

Here, we use high-precision astrometry to search for astrometric
lensing signals and constrain the masses of candidate isolated BHs.
This is the first such attempt made with ground-based adaptive optics.
In \S \ref{sec:mass}, we explain how the lens mass can be estimated
using photometric and astrometric means.  We describe our photometric
selection and observations of three candidate BH microlensing events
in \S \ref{sec:obs} and outline our methods to extract high-precision
astrometry in \S \ref{sec:methods}.  We present photometric and
astrometric microlensing models fitted to the three events in \S
\ref{sec:models}.  Our resulting proper motion fits and lens mass
measurements are detailed in \S \ref{sec:analysis}. In \S
\ref{sec:discussion}, we discuss our findings and in \S
\ref{sec:future} determine the most efficient observing strategies for
detecting the lensing signatures of stellar mass black holes in future
campaigns.  Conclusions are provided in \S \ref{sec:conclusions}.

\section{Estimating the lens mass}
\label{sec:mass}
To confirm that a lens is a black hole, one must constrain the lens
mass to $\gtrsim$5 M$\subsun$ without evidence of a luminous massive
star.  Such attempts have been made by analyzing the light curves of
microlensing events, but with no conclusive success
\citep[e.g.][]{Mao:2002,Bennett02}.   To convey the associated
challenges, we review the methods by which the lens mass can be
estimated.  We consider a lensing event (following the conventions of
microlensing literature, such as those in \citet{Gould14}): The
photometric amplification is given by
\begin{equation}
A = \frac{u^{2} + 2}{u\sqrt{u^{2}  + 4}},
\label{amp}
\end{equation}
where $u$ is the projected source-lens separation in units of the
Einstein radius---the radius of the source image upon perfect
alignment of the observer, source, and lens.  The Einstein radius is
defined in angular units as
\begin{equation}
\theta_{E}=\sqrt{\frac{4GM}{c^{2}}\left( d_{L}^{-1}-d_{S}^{-1} \right)},
\label{eqn:thetaE}
\end{equation}
where $G$ is the gravitational constant, $c$ is the speed of light in
vacuum, $M$ is the lens mass and d$_{L}$ and d$_{s}$ are the lens and
source distances relative to the observer in Euclidean geometry.  
Assuming $d_{L}=$4 kpc and $d_{S}=$8 kpc, a 5 $M\subsun$ BH would
produce an Einstein radius of 2.3 mas.  The time required for the lens
to traverse the Einstein radius is the Einstein crossing time, $t_{E}$
and is recoverable from the light curve.  $t_{E}$ is related to
$\theta_{E}$ via
\begin{equation}
\thetaevec = \murelvec \tE,
\label{eqn:tE}
\end{equation}
where $\murelvec$ is the relative source-lens proper motion.  Notice
that $t_{E}$ scales as $\sqrt{M}$, indicating that  BH lenses produce
statistically longer duration microlensing events than typical stars,
given the same source and lens distances and relative proper motions.
Peak photometric amplification occurs at time $t_0$, when the
source-lens separation reaches its minimum value.  This minimum
projected source-lens separation in units of the Einstein radius is
defined as the impact parameter, $u_{0}$, and is recoverable from the
light curve---amplification grows with decreasing separation.
Omitting parallax effects, the equation of relative motion in units of
the Einstein radius can be written as

\begin{equation}
\vect{u}(t) = \uveco + \tau\thetaEhat,
\label{eqn:uoft}
\end{equation}
where $\tau \equiv \frac{t - t_{0}}{t_{E}}$, and $\thetaEhat$ is a
unit vector in the direction of relative source-lens motion,
perpendicular to $\uveco$.
Earth's annual orbital motion causes deviations in the projected
source-lens separation proportional to the relative parallax,
\begin{equation}
\pi_{\mathrm{rel}} = 1~\mathrm{AU}\left(d_{L}^{-1}-d_{S}^{-1}\right),
\label{eqn:PiRel}
\end{equation}
which impacts the magnified photometric signal
\citep{Gould:1992,Gould:1994}. If the timescale of the lensing event
is long enough, the light curve may show a detectable asymmetry as a
result, which can be used to determine the ``microlensing parallax''
in units of the Einstein radius,
\begin{equation}
\vec{\boldsymbol{\pi}}_E \equiv \frac{\pi_{\mathrm{rel}}}{\theta_E}\thetaEhat.
\end{equation}
The detectability of parallax effects increases with lens mass since
the projected source-lens separation, $\vect{u}(t)$, scales as at
least $\pi_E t_E^2$ and $\pi_E t_E^3$ for the components of the
parallax that are parallel and perpendicular to the source-lens
motion, respectively \citep{Smith03,Gould04}. If the Einstein radius
can be measured, one can then recover $\pi_{\mathrm{rel}}$ and use
Equations \ref{eqn:thetaE} and \ref{eqn:PiRel} to solve for the lens
mass,

\begin{equation}
M = \frac{\theta_{E}}{\kappa\pi_E},
\label{eqn:mass}
\end{equation}
where $\kappa \equiv \frac{4G}{1~AU\cdot c^{2}}$.

\begin{figure*}
\centering
\subfigure{
\includegraphics[width=8cm]{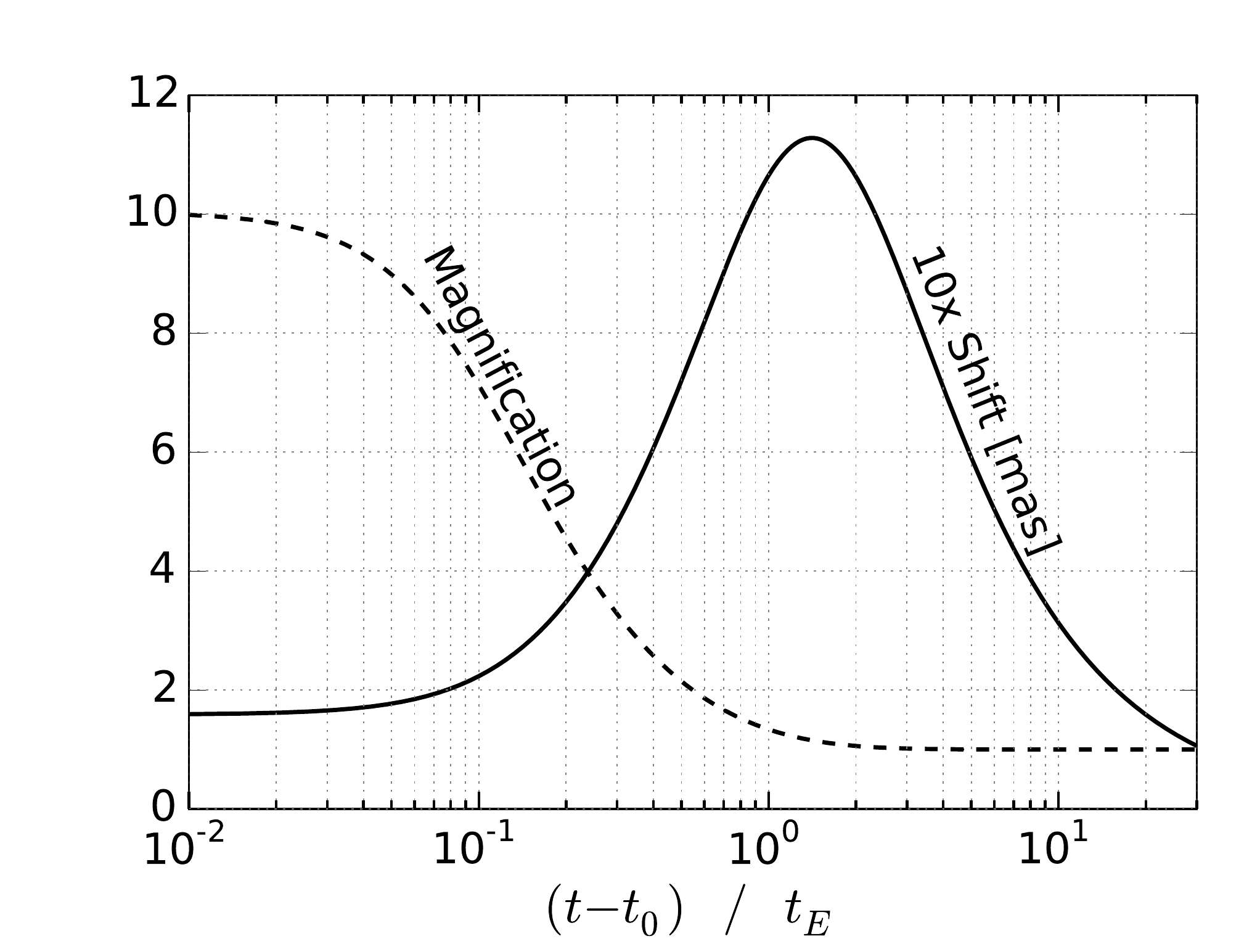}
}
\subfigure{
\includegraphics[width=8cm]{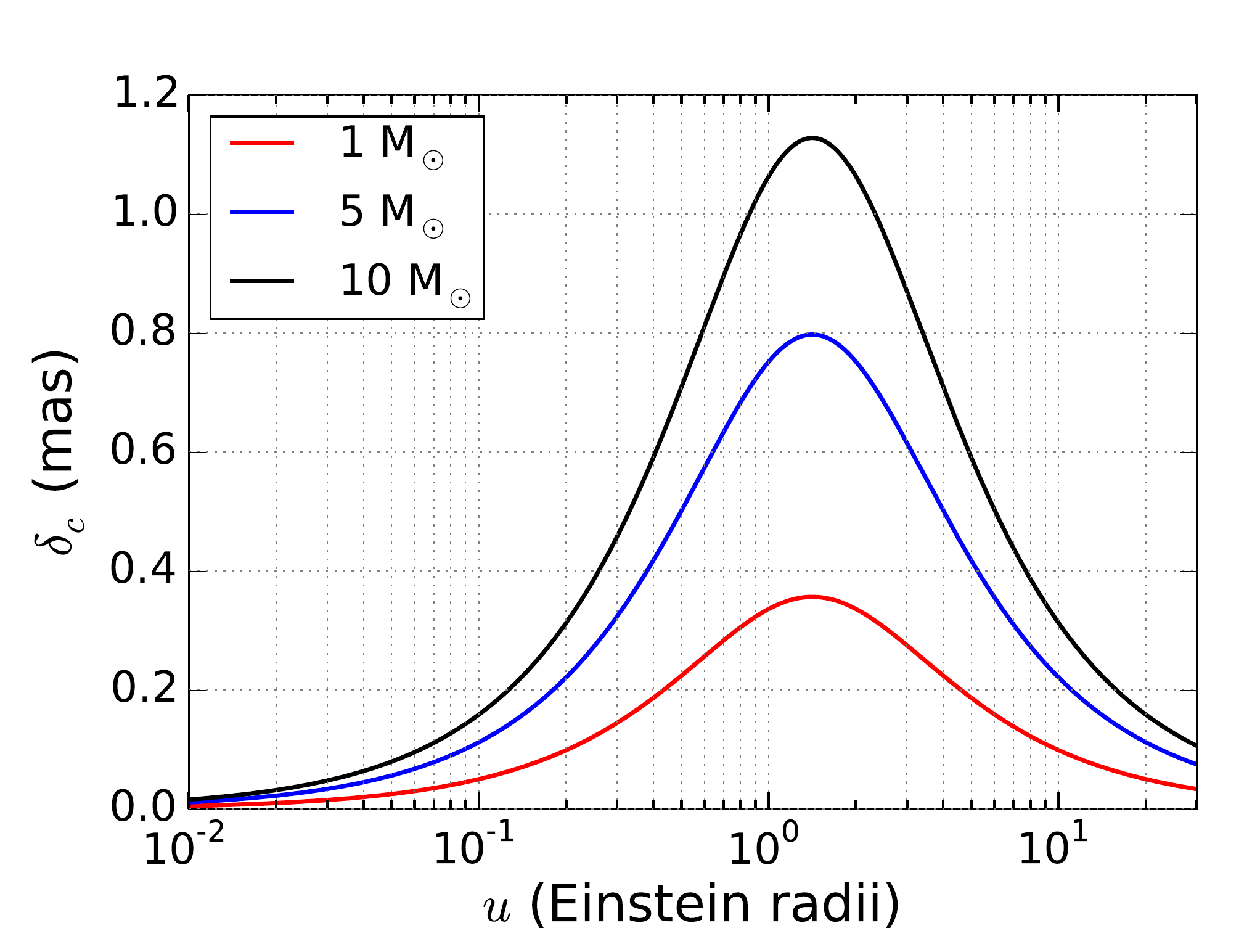}
}
\caption{\textit{Left}: A characteristic example of the photometric
  magnification (dashed line) and astrometric shift (solid line) of
  the lensed source as a function of time since closest approach,
  normalized to Einstein units.   Source magnification is greatest at
  minimum separation, while the astrometric shift reaches a maximum at
  $\left(t-\tnaught\right)/\tE=\sqrt{2}$.  The curves are calculated
  assuming a 10 M$\subsun$ lens at a distance of 4 kpc from Earth, a
  source distance of 8 kpc from Earth, and impact parameter $\uo$=0.1.
  \textit{Right}:  The astrometric shift in the position of a lensed
  source as a function of the projected separation between the star
  and the lens, $u$, in units of Einstein radii.  The curves are
  calculated assuming a distance of 8 kpc between the Earth and the
  star, and a distance of 4 kpc between the Earth and the lens. The
  three different curves are for lens masses of 1 M$\subsun$, 5
  M$\subsun$, and 10 M$\subsun$. The horizontal axis can be converted
  to units of time using the Einstein-radius crossing time. For the 10
  M$\subsun$ case, the Einstein radius is $\sim$ 4 mas and the
  crossing time is typically $>$100 days.}

\label{fig:signals}
\end{figure*}

In principle, if the Einstein ring is resolvable by interferometric
means, $\theta_{E}$ can be measured \citep{Delplancke01}, but this has
yet to be done. For a certain subset of microlensing events, $\thetaE$
can be measured from photometry alone using finite source effects,
which occur when the source approaches or crosses over a caustic.
If the lens is a single object, finite source effects are rarely
measurable, requiring a nearly direct passage of the source over the
lens \citep{Witt:1994,Gould:1994b}.   Even in these fortuitous cases,
constraint of $\thetaE$ is limited to the precision to which both the
microlens parallax and the angular diameter of the source can be
measured from photometry and/or spectroscopy \citep[e.g][]{Alcock97,
  Albrow00, Yoo04, Gould09, Zub11, Choi12}, and the 
most precise mass constraint reported in the literature is 
$\sim$15\%-20\% \citep[e.g.][]{Yee:2009}.
For binary lenses, $\thetaE$ is more routinely measured through
crossings or close approaches of caustics or cusps.
Alternatively, if the lens is luminous and can be viewed at large
enough separation from the source after the event, $\murelvec$ can be
measured and $\thetaE$ can be determined from Equation \ref{eqn:tE}.
In practice, this technique has rarely been used as it requires high
resolution imaging, high lens-source relative proper motions
\citep[e.g.][]{Alcock:2001,An02,Kozlowski:2007,Batista:2015} and is
not applicable to faint or non-luminous lenses (i.e. BHs).

Previously, a number of BH candidates have been proposed based on the
combination of microlensing parallax measurements with Galactic models
that place statistical constraints on $\murel$, and thus on $\thetaE$
via Equation \ref{eqn:tE} \citep{Alcock95, Mao:2002, Bennett02,
  Poindexter05, Dong07, Shvartzvald:2015, Wyrzykowski:2015, Yee15}.
Microlensing parallax measurements are subject to a fourfold
degeneracy that arises because the light curve does not distinguish
the side on which the lens passes the source
\citep[i.e. $\pm u_0$][]{Smith03} and the jerk-parallax degeneracy
\citep{Gould04}.

Although the interpretation of microlensing parallax measurements with
Galactic models can help to infer ensemble properties of lenses
(e.g. cumulative mass and distance distributions), it yields only weak
constraints on the lens mass for any single event.  It is especially
problematic for BH lenses, which might have different spatial and
dynamical distributions than stars due to factors such as supernova
birth kicks.   Alternative approaches are needed to make the first
robust detection of an isolated BH.

Astrometric measurements of the lensed source provide a direct measure
of $\thetaE$ and $\murelvec$ and thus can be used to overcome the
microlensing parallax degeneracies and dependences on Galactic models
that have plagued photometric attempts to constrain lens masses.  The
potential of this technique has been known and studied for over a
decade \citep[e.g.][]{Miyamoto:1995, Hog:1995, Walker:1995,
  Paczynski:1998, Boden:1998, Han99, Jeong99, Gould14}.  During a
microlensing event, the images are unresolved, but the center of light
is shifted relative to the true position of the source by

\begin{equation}
\deltavec(t) = \frac{\theta_{E}}{u(t)^{2}+2}\uvec(t),
\label{eqn:astroshift1}
\end{equation}
\citep{Walker:1995}.  Combining Equations  \ref{eqn:uoft} and \ref{eqn:astroshift1} yields

\begin{equation}
\deltavec(t) = \frac{\theta_{E}}{\tau^{2} + u_{0}^{2} +2} \left[\uveco + \tau\thetaEhat\right].
\label{eqn:astroshift2}
\end{equation}

Figure \ref{fig:signals} shows an example of both the photometric and
astrometric signal induced as a function of the projected source-lens
separation, $u$.  Note that the photometric peak occurs at minimum
separation, $u = u_{0}$ (at t = $t_{0}$), whereas the maximum
astrometric shift occurs at $u = \sqrt{2}$.   Typical astrometric
shifts, even those induced by $\sim$5 $M_\odot$ black holes, are
sub-milliarcsecond (mas) in scale (Figure \ref{fig:signals}).
Detections require the high astrometric precision of facilities like
the Keck adaptive optics system feeding the NIRC2 instrument.
Previous NIRC2 studies have demonstrated astrometric precisions as low
as $\sim0.15$ mas \citep{Lu10}.
Here we use Keck/NIRC2 to make the first ground-based attempt to detect isolated BHs.

%

\begin{figure*}
\begin{center}
\subfigure{%
\includegraphics[width=5cm]{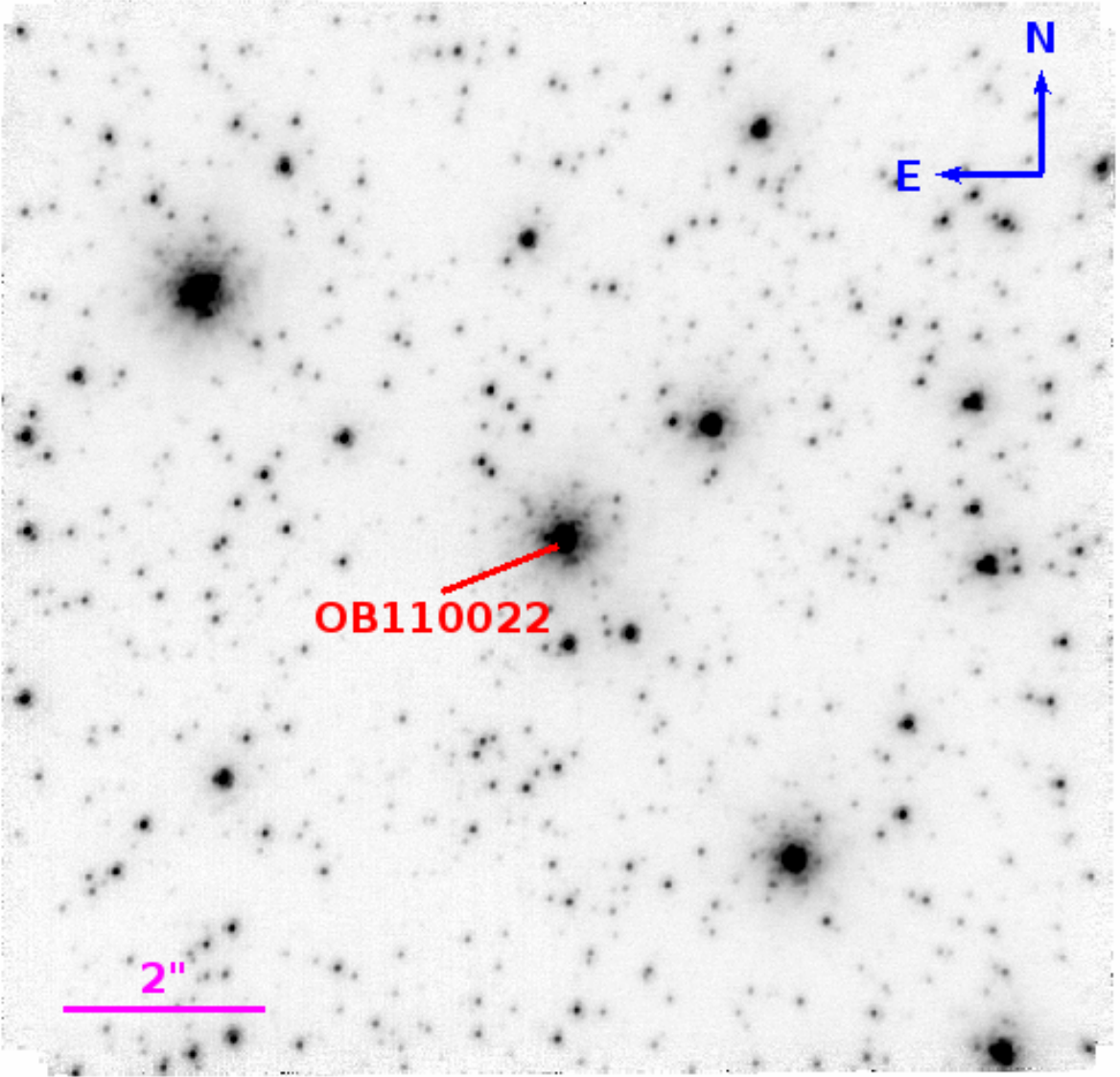}
}
\subfigure{%
\includegraphics[width=5cm]{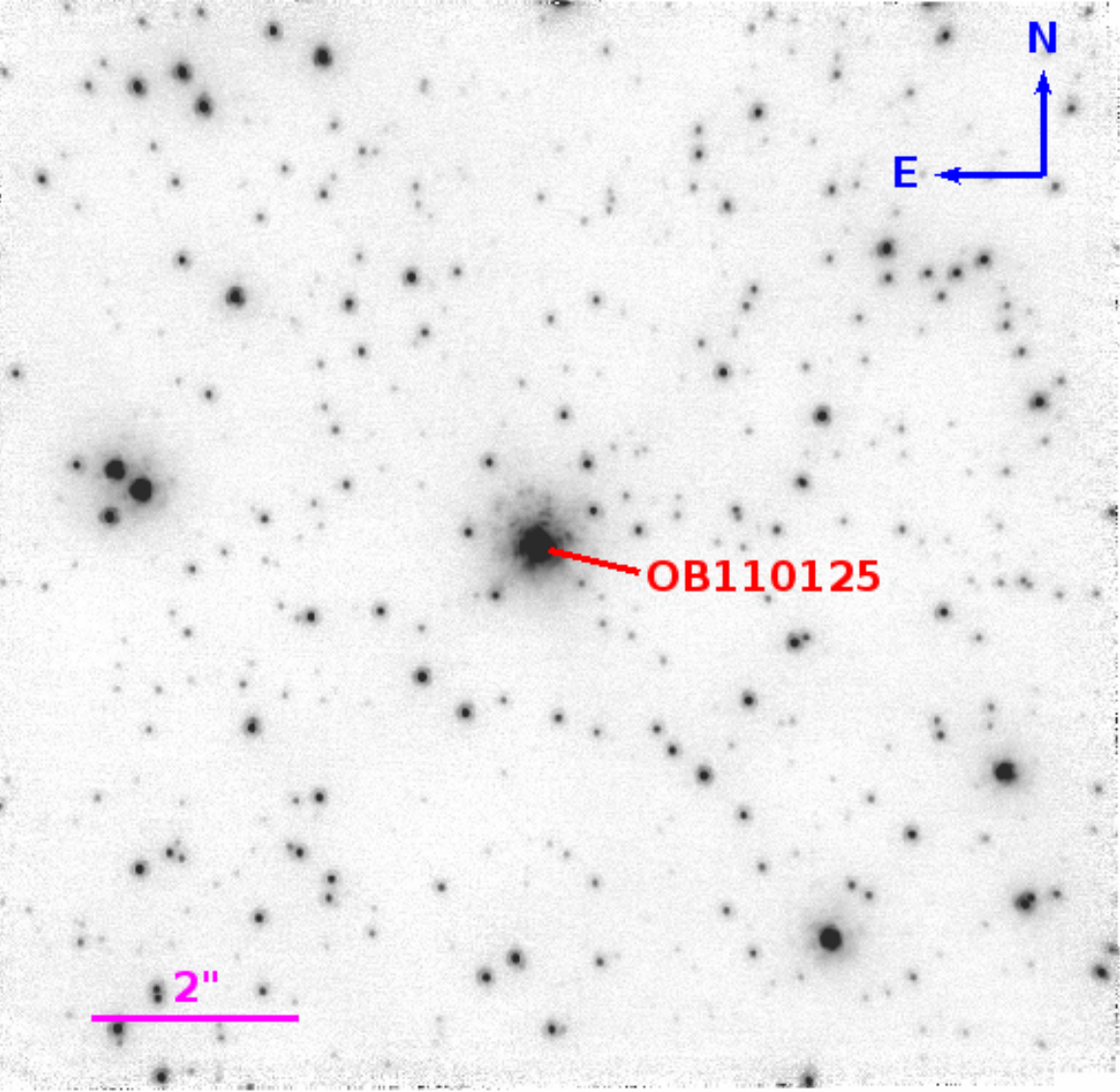}
}
\subfigure{%
\includegraphics[width=5cm]{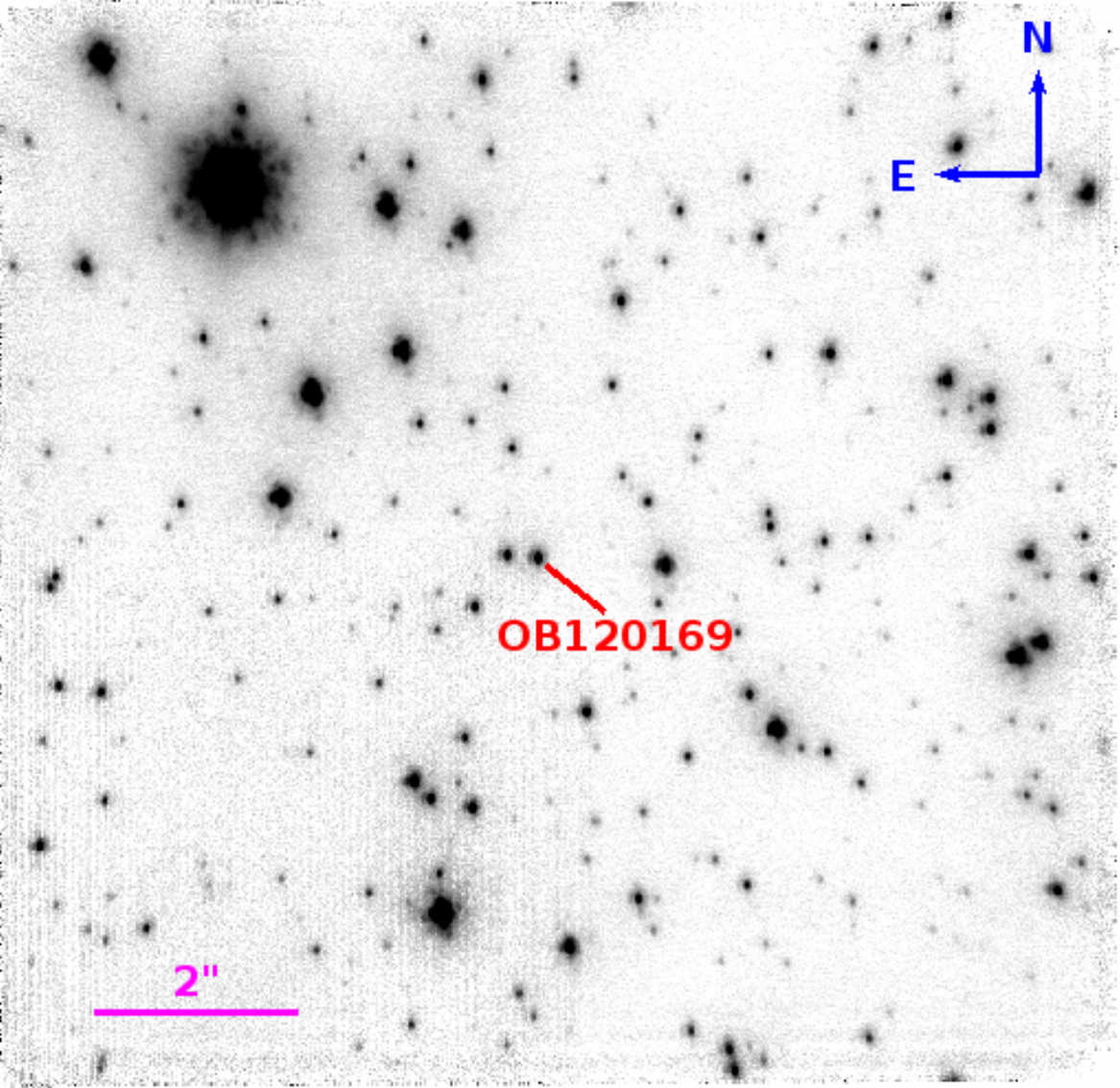}
}
\caption{The three observed 10\arcsec x 10\arcsec fields centered on the targets of interest:  OB110022 (left panel), OB110125 (middle panel), and OB120169 (right panel).}

\end{center}
\label{fig:fields}

\end{figure*}

\section{Observations}

\label{sec:obs}

\subsection{Photometry from the OGLE survey}

We use photometry from the Optical Gravitational Lensing Experiment
survey \citep[OGLE,][]{Szymanski00}. OGLE is a continuous, long term
survey carried out with the 1.3-m Warsaw telescope at the Las Campanas
Observatory in Chile. The survey is currently in its fourth phase
(OGLE-IV), with the telescope equipped with a 32-CCD mosaic camera,
and focuses on monitoring stars toward the Galactic bulge for
microlensing.
See \citet{Udalski15b} for more details on the project. Currently, the OGLE survey
discovers, in real time, over 2000 microlensing events per year with its
Early Warning System \citep{Udalski03}\footnote{\url{http://ogle.astrouw.edu.pl/ogle4/ews/ews.html}}.
The $I$-band light curves used in this study come from an independent
off-line reduction, optimized for these events and using an improved
lens position, which used the OGLE photometric
pipeline and Difference Image Analysis (DIA) package
\citep{Wozniak00}.

Prior to the 2011 and 2012 astrometric observing seasons, we
identified the longest duration events with reasonable magnification
($A_{\rm max} \gtrsim10$) using the OGLE real-time Early-Warning
System \citep{Udalski03}.  In addition, we required that the sources
showed no hints of blending and had baseline magnitudes $I\lesssim19$.
We ultimately selected the following three events with long timescales
($t_{\rm E} > 130 {\rm \, days}$) for astrometric follow-up described
in \S\ref{sec:astrometryobs}:  OGLE-2011-BLG-0022,
OGLE-2011-BLG-0125, and OGLE-2012-BLG-0169 (hereafter OB110022,
OB110125, and OB120169).    Since the astrometric monitoring
had to start near their peaks, the selections could not be done using
the full light
curves but were instead performed based on modeling
the rising part of the light curves prior to their peaks.
For two events, OB110022 and OB110125, the light curves ultimately
showed significant asymmetry with respect to the peak. The single-lens
modeling prior to peak over-estimated the event time scale for both
events and binary-lens models are favored.
Modeling of the complete photometric dataset is presented in \S \ref{sec:photmodel}.

%
\begin{deluxetable*}{lrrrrrrrrr}
\tabletypesize{\footnotesize}
\tablewidth{0pt}
\tablecaption{AO Observations}
\tablehead{
Event  & RA (J2000) & Dec (J2000) & Date & $N_{\mathrm{exp}}$ & $N_{\star}$  & 
Strehl &  FWHM & $\sigma_{\mathrm{pos}}$ & $\sigma_{\mathrm{aln}}$ \\ 
& [hr] & [deg] & [UT] & & & & [mas] & [mas] & [mas]
}
\startdata
\\
OB110022 & 17:53:17.93 & -30:02:29.3 
    & May 25, 2011    & 27 & 285 & 0.14 &  91 & 0.56 & 0.24 \\ 
& & & July 7, 2011    & 16 & 178 & 0.13 &  69 & 0.66 & 0.34 \\ 
& & & June 23, 2012   & 40 & 701 & 0.24 &  70 & 0.31 & 0.21 \\ 
& & & July 10, 2012   & 34 & 717 & 0.26 &  68 & 0.25 & 0.18 \\ 
& & & April 30, 2013  & 22 & 485 & 0.24 &  71 & 0.30 & 0.00 \\ 
& & & July 15, 2013   & 30 & 636 & 0.34 &  60 & 0.20 & 0.21 \\ 
\\ 
OB110125 & 18:03:32.95 & -29:49:43.0 
    & May 23, 2012    & 21 & 104 & 0.10 &  96 & 0.39 & 0.63 \\ 
& & & June 23, 2012   & 33 & 327 & 0.36 &  57 & 0.09 & 0.21 \\ 
& & & July 10, 2012   & 18 & 221 & 0.21 &  70 & 0.25 & 0.23 \\ 
& & & April 30, 2013  & 48 & 332 & 0.29 &  64 & 0.13 & 0.00 \\ 
& & & July 15, 2013   & 39 & 329 & 0.36 &  57 & 0.12 & 0.14 \\ 
\\ 
OB120169 & 17:49:51.38 & -35:22:28.0 
    & May 23, 2012    &  5 &  35 & 0.10 & 110 & 0.84 & 1.19 \\ 
& & & June 23, 2012   & 10 & 122 & 0.24 &  69 & 0.38 & 0.27 \\ 
& & & July 10, 2012   & 22 & 192 & 0.29 &  64 & 0.22 & 0.30 \\ 
& & & April 30, 2013  & 31 & 207 & 0.29 &  61 & 0.17 & 0.00 \\ 
& & & July 15, 2013   & 11 &  84 & 0.26 &  74 & 0.47 & 0.38 \\ 
\enddata
\tablenotetext{}{$N_{\star}$: Number of stars detected. Strehl and
FWHM are the average values over all individual
exposures. $\sigma_{\mathrm{pos}}$ and $\sigma_{\mathrm{aln}}$ are
calculated after cross-epoch transformation from the median 
of all stars with r$<$4 and Kp$<$19 mag. 
}
\label{tb:AOobs}
\end{deluxetable*}

\subsection{Astrometry with Keck/NIRC2}
\label{sec:astrometryobs}
Astrometric measurements in this study derive from multi-epoch imaging
observations of three 10\arcsec x 10\arcsec fields towards the
Galactic Bulge from the W.~M. Keck II 10 m telescope. Images were
obtained with the near infrared camera, NIRC2 (PI: K. Matthews) behind
the laser guide star adaptive optics facility
\citep{Wizinowich:2006,vanDam:2006}.  Figure \ref{fig:fields} shows
each of the three observed fields, which are centered on OB110022,
OB110125, and OB120169 respectively.

Table \ref{tb:AOobs} provides a summary of astrometric observations.
Each target was observed in 5--6 epochs spanning 1--2 years.  All
epochs correspond to the post-peak phase of the microlensing event
during which the magnification declines; but the astrometric shift
reaches a maximum at $t = t_0 + t_E \sqrt{2 - u_0^2}$ for a
single-lens event (see Figure \ref{fig:signals}).

To aid reconstruction of bad pixels in NIRC2 and avoid detector
persistence from bright sources, the telescope was dithered in a
continuous random dither pattern with three images taken at each
position, with individual integration times of 30 s per frame, split
into 10 co-adds $\times$ 3 s in the Kp filter ($1.95-2.30 \mu$m) to
avoid saturation of the brightest sources. The dither pattern was
confined to a small 0\farcs7 $\times$ 0\farcs7 box to minimize
astrometric errors due to residual distortions.  The AO/NIRC2
distortion solution is accurate to the $\sim$ 1 mas level
\citep{Yelda10}.   For details on this correction we refer the reader
to \citet{Yelda10}.   The same study derives a plate scale of 9.952
$\pm$ 0.002 mas/pix, which we adopt for our analysis. The visual
seeing during May 2011, July 2011, and May 2012 observations was $\sim
2\arcsec$, much worse than average for Mauna Kea conditions, yielding
larger astrometric uncertainties in these epochs.

\section{Astrometric Data Analysis}
\label{sec:methods}

\subsection{Raw reduction \& generation of stellar catalog}

Initial reduction of raw data from each epoch of observation was
carried out with our custom NIRC2 reduction pipeline
\citep{Stolte:2008,Lu09} and included flat field and dark calibration,
sky subtraction and cosmic ray removal.  Corrections for distortion,
achromatic differential atmospheric refraction (DAR), and image shifts were applied
using the IRAF routine, \texttt{Drizzle} \citep{Fruchter02}, as
described in \citet{Yelda10}.
We note that chromatic DAR effects are small given that all are
targets were observed at approximately the same zenith angle in every epoch
\citep{GublerTytler98}.
The individual cleaned exposures were co-added, weighted by strehl, to
produce a combined image.  This summation was restricted to individual
frames displaying core FWHM $< 1.25 ~\mathrm{FWHM_{min}} $, where
$\mathrm{FWHM_{min}}$ is the minimum FWHM of all frames of the
particular field and epoch. May 2012 observations of OB110022 were
discarded because of poor data quality.

On each combined map, we used the point-spread function (PSF) fitting
routine, Starfinder \citep{Diolaiti00} to extract a stellar catalog of
spatial coordinates and relative brightness.   First,  the PSFs of a
subset of stars (hereafter ``PSF stars") were averaged to derive a
mean PSF. The PSF is then cross-correlated with the image and stars
were identified as peaks with a correlation above 0.8.  To achieve
precise centroiding, we locally fit the mean PSF to each identified
star.  The single-epoch astrometric precision is thus very sensitive
to the derivation of the mean PSF and so strict criteria were applied
in the selection of PSF stars.  Specifically, stars contributed to the
mean PSF derivation if they were bright (typically Kp $<$ 18 mag),
isolated, and within 4\arcsec of the center of the field.  The latter
criterion avoids detector edge effects and ensures all PSF stars are
close to the target of interest, mitigating astrometric error
attributed to spatial variation of the PSF caused by instrumental
aberrations and atmospheric anisoplanatism.   For OB110022
observations in May and July 2011, the PSF showed significant
variation across the field and thus a stricter radial cut of
$2.5\arcsec$ from the target was applied.

The resulting starlists produced by Starfinder contain positions and
fluxes for each star in detector units of pixels and counts,
respectively.
Fluxes are calibrated using 2MASS K-band magnitudes
\citep{Skrutskie062MASS} of several bright stars within the field of view
and the average brightness for each target is reported in Table
\ref{tb:AOobs}.

\begin{figure*}
\centering
\includegraphics[scale=0.5]{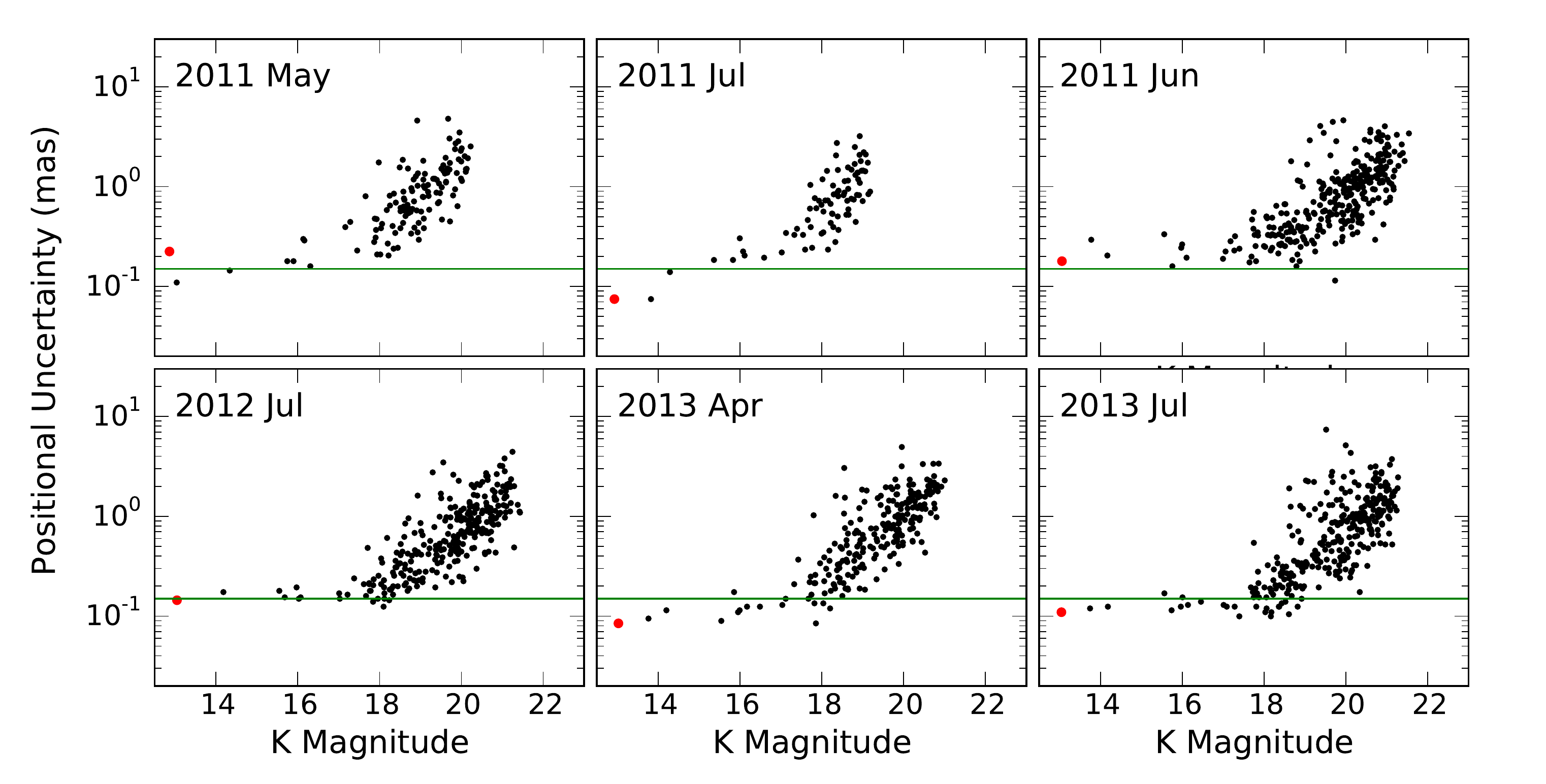}
\caption{The centroiding error versus K-band brightness for stars
  ({\em black points}) within 4$\arcsec$ of target OB110022 ({\em red
    point}).  Each epoch is presented in a different
  panel. Centroiding errors are the standard deviation of the star's
  mean position in 3 submaps, each composed of the co-addition of a
  different third of the images. The {\em green lines} indicate the
  desired astrometric precision of 0.15 mas, which is achieved in half
  of the epochs for K $\lesssim$ 18 mag.}
\label{fig:MedPosErr022}
\end{figure*}

\begin{figure*}
\centering
\includegraphics[scale=0.5]{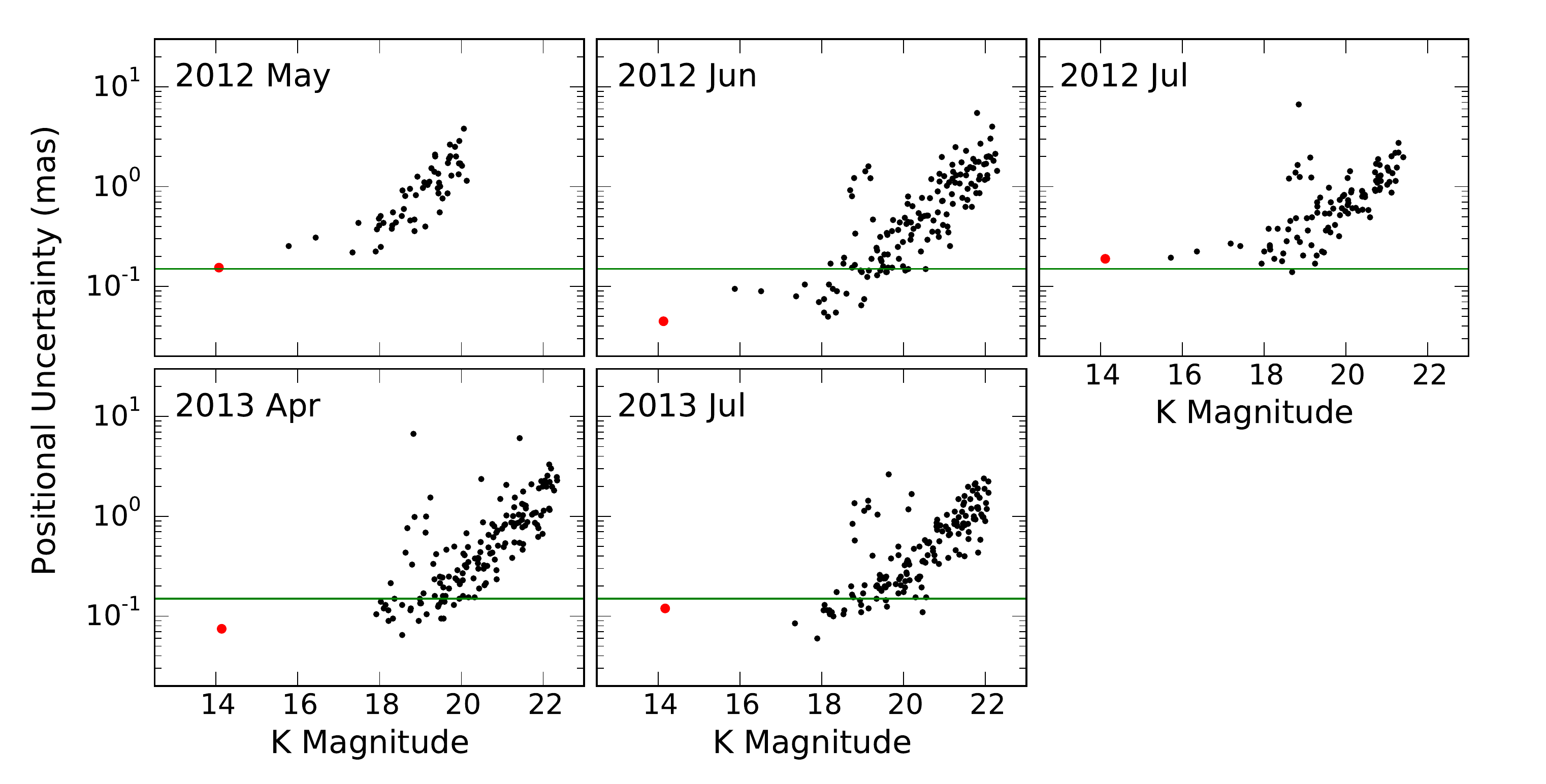}
\caption{The centroiding error versus K-band brightness for stars
  ({\em black points}) within 4$\arcsec$ of target OB110125 ({\em red
    point}).  Each epoch is presented in a different
  panel. Centroiding errors are the standard deviation of the star's
  mean position in 3 submaps, each composed of the co-addition of a
  different third of the images. The {\em green lines} indicates the
  desired astrometric precision of 0.15 mas, which is achieved in most
  epochs for K $\lesssim$ 18 mag.}
\label{fig:MedPosErr125}
\end{figure*}

\begin{figure*}
\centering
\includegraphics[scale=0.5]{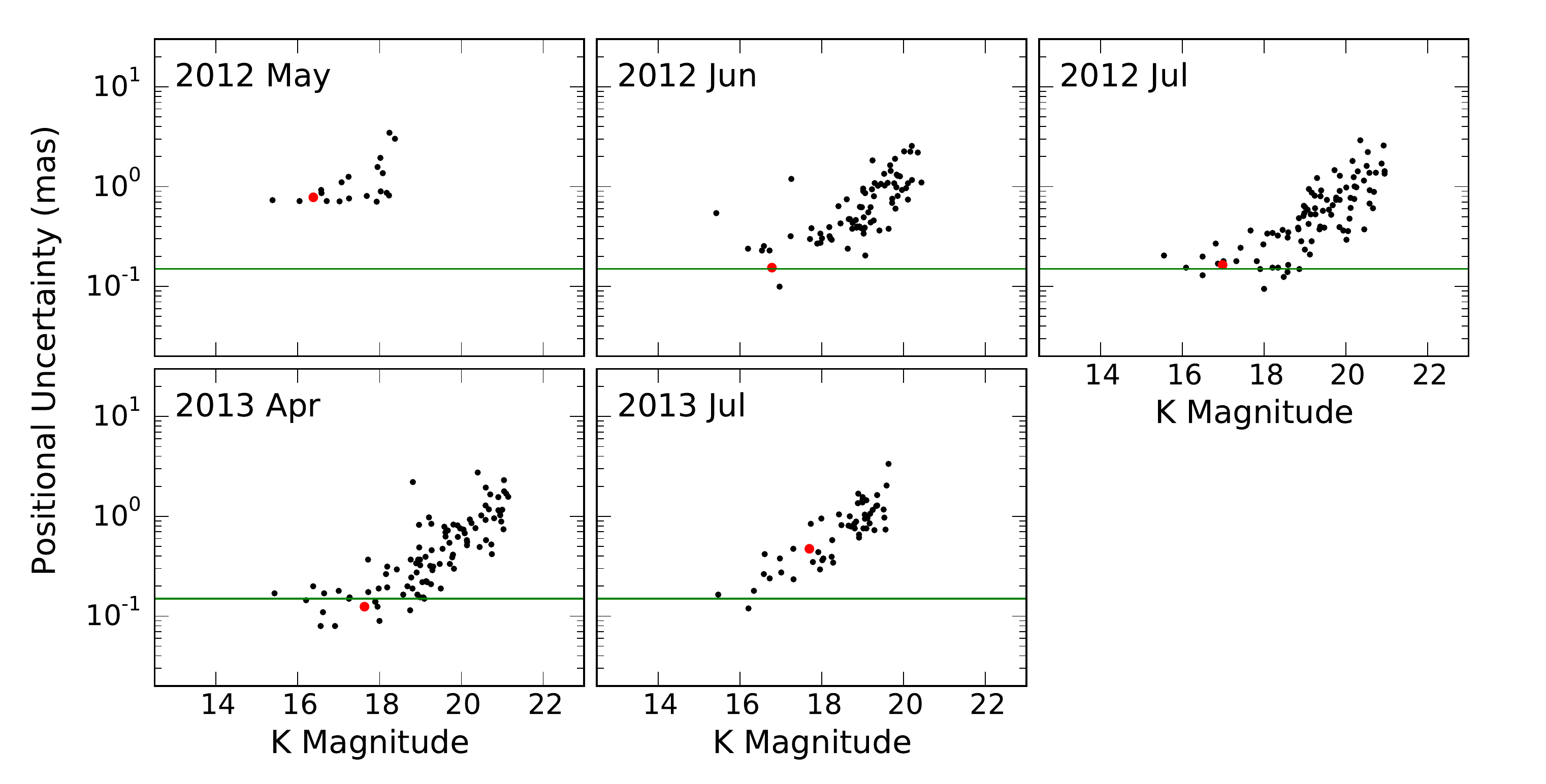}
\caption{The centroiding error versus K-band brightness for stars
  ({\em black points}) within 4$\arcsec$ of target OB120169 ({\em red
    point}).  Each epoch is presented in a different
  panel. Centroiding errors are the standard deviation of the star's
  mean position in 3 submaps, each composed of the co-addition of a
  different third of the images. The {\em green lines} indicates the
  desired astrometric precision of 0.15 mas, which is achieved in most
  epochs for K $\lesssim$ 17 mag.}
\label{fig:MedPosErr169}
\end{figure*}

We derive centroiding errors, $\sigma_{pos}$, for each star in each epoch.   
The centroiding error adopted for each
star was the standard error on the mean of its position in three ``sub
maps".  Each of these sub maps was produced by shifting and co-adding 
a different third of the frames selected from a Strehl-ranked list such that all
sub maps had similar Strehl.  A linear translation was then applied
to each map, minimizing the astrometric scatter between each sub map
and the combined map.   The transformation was derived using stars
brighter than a cutoff magnitude, $K_{\mathrm{cut}}$.   The relative
effectiveness of using different $K_{\mathrm{cut}}$ values results
from a tradeoff between number of stars and their average brightness.
Three values, $K_{\mathrm{cut}} =$ 18, 19 and 20 mag,  were tested for
each field producing only minor differences in centroiding error.  The
best precision was achieved using $K_{\mathrm{cut}}$ = 18 mag for the
OB110022 field, and 20 for the OB110125, and OB120169 fields.
Figures \ref{fig:MedPosErr022}-\ref{fig:MedPosErr169} show the
resulting positional errors, $\sigma_{pos}$ for all stars in each epoch of
observation as a function of brightness.  In most epochs, the target
astrometric precision of 0.15 mas (red line) is achieved for K
$\lesssim$ 18 mag.

%

%

%

\subsection{Cross-epoch Alignment \& Proper Motion Fitting }

The first step in computing proper motions is to align images from all
epochs into a common coordinate system.   Alignment is often
complicated by epoch-dependent systematic effects, which can
ultimately dominate the proper motion uncertainties.  More
specifically, changes in instrument conditions and performance
(e.g.~PSF, pixel scale) from epoch to epoch cause significant
variations in relative astrometry.
We correct for these systematic effects by transforming combined
images from all epochs into the pixel coordinate system of the April
2013 observations using $\chi^2$-minimization to find the best-fit 
linear transformation of spatial coordinates
(x, y):
\begin{equation}
x^{\prime} = a_0 + a_1x + a_2y + a_3x^2 + ...
\label{eqn:transx}
\end{equation}
\begin{equation}
y^{\prime} = b_0 + b_1y + b_2x + b_3y^2 + ...
\label{eqn:transy}
\end{equation}
where $a_i$ and $b_i$ are fitted
coefficients.    This first three terms give a linear transformation
independent in $x$ and $y$,
effectively correcting for differences in rotation, shear, pixel
scale, and translation between star lists from different epochs based
on the change in the positions of individual stars. Higher-order 
transformations are tested and used, when necessary, as discussed
further below. 
The transformation is first performed assuming all stars have zero
motion. Proper motions are then derived for the stars and the
transformation process is repeated now accounting for the motion of
each star. This iterative procedure of deriving the coordinate
transformations and stellar proper motions ultimately produces precise
relative proper motions in a reference frame that is at rest with
respect to the mean motion of all stars within the field. 

Note that we have implicitly assumed that the reference stars have
negligible parallax. This assumption is supported given that simulated
stellar populations using the TRILEGAL galaxy model \citep{Girardi12}
in the direction of the targets shows that our sample of reference
stars is likely composed of $>$83\% bulge stars and $>$95\% of stars
with distances exceeding 5 kpc. Also, the handful of reference stars
that can be matched
with OGLE sources at I-band, given the limited OGLE spatial
resolution, have I$-$Kp colors consistent with
highly extincted, thus distant, bulge stars. 

The quality of the cross-epoch alignment depends on several factors,
including which stars are used to find the best-fit transformation,
how these stars are weighted, and the order of the transformation
equations.
The sample of stars used to derive the transformation was restricted
to those detected in all epochs and within 4\arcsec of the
microlensing targets.
The microlensing targets were omitted since their motion would not be
linear if they are lensed. 
We further restricted the astrometric
reference stars to bright stars and we tested brightness cuts of
$K_{\mathrm{cut}}=$ 18, 20, and 22 mag.  
Extending the reference star sample from 
$K_{\mathrm{cut}}=$ 18$\rightarrow$20 significantly improves the
cross-epoch alignment since the number of reference stars increases
from 0-100 to $>$400. Some epochs are sufficiently deep that 
extending from $K_{\mathrm{cut}}=$ 20$\rightarrow$22 added $>$100 
more stars to the sample. Therefore, we adopted $K_{\mathrm{cut}}=$ 22
for all of the targets and epochs. The astrometric reference
stars were weighted based on positional uncertainties, as outlined in
Appendix \ref{sec:weighting}; and we find that the proper motion fits
are independent of our choice of weighting scheme.  
We also tested different orders (O=1,2,3) of polynomial
transformations between each epoch using an F-test (Appendix
\ref{sec:align_order}). OB110022 shows minimal
improvement when advancing to 2nd order,
therefore we adopt O$=$1 for this source. 
OB110125 and OB120169 showed improvement going from O$=1\rightarrow2$,
therefore we adopted O$=$2 for these sources.
We note that microlensing fits were ultimately run for both O$=$1 and
O$=$2 for both OB110022 and OB120169 and there
was negligable change in the final lens-mass posteriors.

In addition to the random error on the position of each star,
$\sigma_{pos}$, there is
additional error due to uncertainties in the coordinate frame
alignment. To capture this alignment error, a half-sample bootstrap
resampling analysis is performed to derive the transformation
parameters using different half-sample sets of stars
\citep{Babu:1996}. The alignment error, $\sigma_{aln}$, for each star
is taken as the standard deviation
of the transformed positions from
all the samples
\citep[see][for further
justification]{Ghez:2000,Ghez:2005,Clarkson:2012}.
The alignment error is typically smaller than the average
centroiding error for all stars and is independent of brightness.
However, the alignment errors produce comparable uncertainties for the O=2
transformations that we have adopted for OB110125 and OB120169, likely
due to the small number of bright stars that dominate the alignment
and systematic errors from PSF variations over the field of view. 
Table \ref{tb:AOobs} presents the median positional and
alignment error for each epoch for all stars with r$<$4'' and
Kp$<$19 mag. 
The final positional error adopted for each lensing target is
$\sigma = \sqrt{\sigma_{pos}^2 + \sigma_{aln}^2}$ and is, 
on average, 
%
%
$\sim$0.26 mas for OB110022, 
$\sim$0.34 mas for OB110125, and 
$\sim$0.68 mas for OB120169 for each epoch.

\begin{figure}
\centering
\includegraphics[width=8cm]{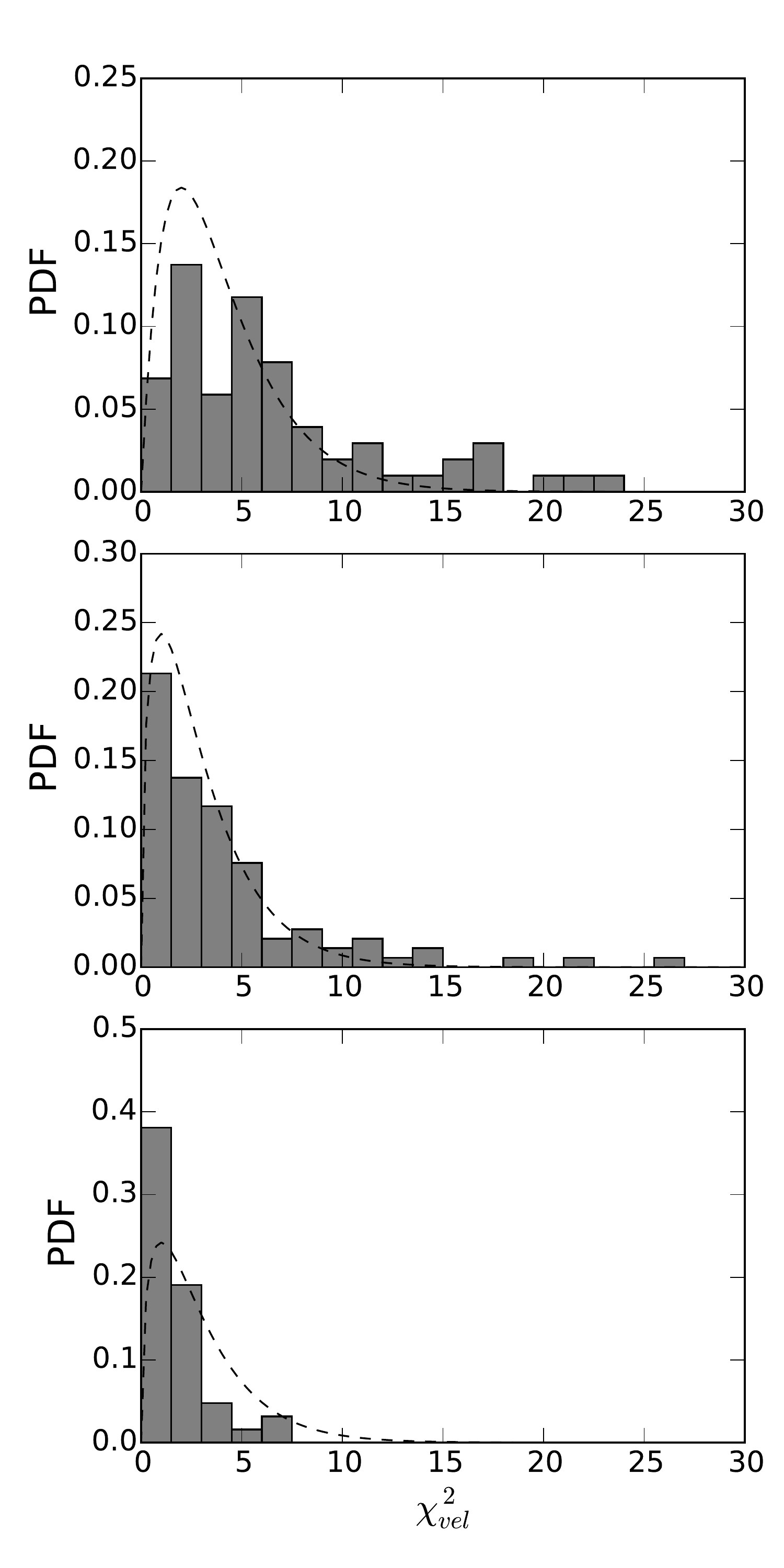}
\caption{Distribution of $\chi^2_{\mathrm{vel}}$ for all stars around
OB110022 ({\em left}), OB110125 ({\em center}), and OB120169 ({\em right}).}
\label{fig:hist_chi2}
\end{figure}

Proper motions are fit for all stars in the field of view. 
To evaluate the quality of the final cross-epoch transformations, we
examine the distribution of $\chi^2_{vel}$ values from the proper motion fits
for all of the unlensed stars detected in all epochs (Figure
\ref{fig:hist_chi2}). These $\chi^2_{vel}$ values are calculated from
the residuals from the proper motion fits for each star, weighted by
the total uncertainty, $\sigma$, which includes both positional and
alignment errors. The distribution of $\chi^2_{vel}$
for all the stars is expected to follow a standard $\chi^2$
distribution with $2*(N_{\mathrm{epochs}} - 2)$ degrees of freedom
(2 free parameters in the velocity fit in each direction), 
which is also shown in Figure \ref{fig:hist_chi2}. 
Deviations from the expected distribution are likely due to systematic
errors (e.g.~spatial PSF variations) that cannot be captured with
simple error re-scaling techniques. Reducing or accounting for these 
systematic errors is the subject of on-going adaptive optics PSF
modeling projects \citep[e.g.][]{Britton:2006,Fitzgerald:2012} and is beyond the
scope of this paper. 

For each lensing target, the final 
positions and uncertainties at each epoch used for both the proper
motion and microlens fits are presented in Tables
\ref{tab:obs_ob110022}, \ref{tab:obs_ob110125}, and
\ref{tab:obs_ob120169}. 
All positions are reported as relative offsets
to the mean position with positive values increasing to
the East and the North.

%
\begin{deluxetable}{lrrrcccc}
\tabletypesize{\footnotesize}
\tablecaption{OB110022 Measurements \label{tab:obs_ob110022}}
\tablehead{
MJD & Kp & $\Delta\xpos$ & $\Delta\ypos$ & 
   \multicolumn{2}{c}{$\sigma_{\mathrm{\x}}$} & \multicolumn{2}{c}{$\sigma_{\mathrm{\y}}$} \\
 & & [mas] & [mas] & \multicolumn{2}{c}{[mas]} & \multicolumn{2}{c}{[mas]} \\
 & & & & pos & aln & pos & aln 
}
\startdata
55706.540 &  12.9 &    -5.06 &    -2.96 &  0.29 &  0.19 &  0.16 &  0.13 \\ 
55749.273 &  12.9 &    -3.84 &    -2.55 &  0.09 &  0.26 &  0.06 &  0.23 \\ 
56101.366 &  13.1 &     0.26 &    -0.46 &  0.22 &  0.11 &  0.14 &  0.14 \\ 
56118.168 &  13.0 &     0.35 &     0.06 &  0.12 &  0.08 &  0.17 &  0.15 \\ 
56412.553 &  13.0 &     3.64 &     2.92 &  0.04 &  0.00 &  0.13 &  0.00 \\ 
56488.158 &  13.0 &     4.65 &     2.99 &  0.11 &  0.09 &  0.11 &  0.14 
   \enddata
\tablenotetext{}{}
\end{deluxetable}

%
\begin{deluxetable}{lrrrcccc}
\tabletypesize{\footnotesize}
\tablecaption{OB110125 Measurements \label{tab:obs_ob110125}}
\tablehead{
MJD & Kp & $\Delta\xpos$ & $\Delta\ypos$ & 
   \multicolumn{2}{c}{$\sigma_{\mathrm{\x}}$} & \multicolumn{2}{c}{$\sigma_{\mathrm{\y}}$} \\
 & & [mas] & [mas] & \multicolumn{2}{c}{[mas]} & \multicolumn{2}{c}{[mas]} \\
 & & & & pos & aln & pos & aln 
}
\startdata
56070.686 &  14.1 &    -0.17 &     0.88 &  0.18 &  0.48 &  0.13 &  0.54 \\ 
56101.366 &  14.1 &     0.46 &    -0.09 &  0.05 &  0.24 &  0.04 &  0.18 \\ 
56118.533 &  14.1 &     0.49 &    -0.02 &  0.22 &  0.20 &  0.16 &  0.28 \\ 
56412.553 &  14.1 &    -0.41 &    -0.55 &  0.07 &  0.00 &  0.08 &  0.00 \\ 
56488.523 &  14.2 &    -0.36 &    -0.21 &  0.09 &  0.13 &  0.15 &  0.15 
   \enddata
\tablenotetext{}{}
\end{deluxetable}

%
\begin{deluxetable}{lrrrcccc}
\tabletypesize{\footnotesize}
\tablecaption{OB120169 Measurements \label{tab:obs_ob120169}}
\tablehead{
MJD & Kp & $\Delta\xpos$ & $\Delta\ypos$ & 
   \multicolumn{2}{c}{$\sigma_{\mathrm{\x}}$} & \multicolumn{2}{c}{$\sigma_{\mathrm{\y}}$} \\
 & & [mas] & [mas] & \multicolumn{2}{c}{[mas]} & \multicolumn{2}{c}{[mas]} \\
 & & & & pos & aln & pos & aln 
}
\startdata
56070.686 &  16.4 &     1.01 &    -0.92 &  1.08 &  0.72 &  0.48 &  1.00 \\ 
56101.366 &  16.8 &     0.20 &    -0.32 &  0.11 &  0.26 &  0.20 &  0.24 \\ 
56118.533 &  17.0 &     0.32 &    -0.30 &  0.18 &  0.32 &  0.15 &  0.25 \\ 
56412.553 &  17.6 &    -0.73 &     0.46 &  0.16 &  0.00 &  0.09 &  0.00 \\ 
56488.523 &  17.7 &    -0.81 &     1.07 &  0.07 &  0.25 &  0.88 &  0.34 
   \enddata
\tablenotetext{}{}
\end{deluxetable}

%
%

\section{Fitting Microlensing Models}
\label{sec:models}

\subsection{Photometry-only}
\label{sec:photmodel}

Microlensing models are fit to the OGLE $I$-band light curves of the
three lensed sources using a Markov Chain Monte Carlo (MCMC) analysis.
A standard point-source-point-lens (PSPL) model is first fit to each
event, taking into account microlens parallax effects \citep{Gould04}.
The PSPL model includes seven free parameters, including $\tnaught$,
$\uo$, $\tE$, and the East and North components of the microlensing
parallax vector, $\piEN$ and $\piEE$.  The final two parameters are
the baseline (i.e. unlensed) flux of the source, $\fsrc$, and the flux
from blended sources, $\fb$, that might be in the photometric
aperture.  We consider the twofold degeneracy ($\uo>0$ / $\uo<0$)
\citep{Gould04}, which we denote as $\uoplus$ and $\uominus$ solutions
respectively.

A PSPL model with parallax adequately describes the OB120169 light
curve.  In contrast, fits to the light curves of OB110022 and OB110125
are poor.  These latter light curves have much steeper slopes during
the post-maximum phases compared to the pre-maximum phases.  We find
that in both cases, binary-lens models with microlens parallax provide
substantially better fits to the light curves. 

For the binary-lens model, we introduce three additional parameters,
mass ratio $q$, projected binary separation $s$ in the units of the
Einstein radius, and position angle $\alpha$ between the source
trajectory and the binary axis.  Note that this assumes that the
binary lens orbital motion is negligible.   In a similar fashion to
\citet{Dong07}, we search for the best-fit solutions via an MCMC and
consider the four-fold degeneracy (close/wide binary \citep{Dominik99}
and $\uoplus/\uominus$).

This 10-parameter binary lens model provides a satisfactory fit to the
OB110125 data.  However, the fit to the OB110022 light curve showed
significant residual structure. We considered that these second-order
effects might arise from orbital motion of the binary lens.  To test
this hypothesis, we introduced two additional model parameters, time
derivatives of projected binary separation $\dot{s}/s$ and rotation
angle $\Omega$.  The non-static solution is a significant improvement
over the static solution. 

Given the similarity in their deviations from the PSPL models
(slow rise and steep fall), we consider the possibility that OB110022
and OB110125 are not genuine microlensing events; but, instead, belong
to a previously unknown class of
long-period variable stars.
One unique characteristic of microlensing events is that the magnification is
achromatic (unless there are significant finite-source effects, of which
there is no evidence in the light curves in this study). For variable stars,
the changes in flux are generally associated with variations in color.
We use the sparsely-covered OGLE $V$-band data during the events (with a cadence
of roughly once per 10 days) to check the color variations. With
model-independent linear-regression, the colors do not change at the level
of photometric precision ($1\%$) during the event,
consistent with the microlensing interpretation.

\subsection{Astrometric model with photometric priors}

We fit an astrometric lensing model to our NIRC2 data in order to measure lens masses.
We adopt simple point-source, point-lens astrometry models for all
three cases. In principle, a binary-lens model should be adopted for
OB110022 and OB110125. However, in the limit where the observations
are consistent with linear proper motion and no astrometric lensing
signal is detected, which is the case for all three of our targets
(see \S\ref{sec:propermotions}), the PSPL should be sufficient for
estimating mass constraints. We also ignore parallax effects when
modeling the astrometry.
As shown by \citet{Boden:1998} and \citet{Han:2000}, microlens
parallax has negligible effect on the astrometric microlensing signals
given our astrometric precision. For the final best-fit PSPL
astrometric models, we confirm that parallax effects shift the
expected astrometric positions typically by less than 0.10 mas and
always less than 0.2 mas, within our measurement uncertainty.

We employ a Bayesian inference method to model the
astrometry. Posterior probability distributions derived from the
photometric MCMC analysis (\S \ref{sec:photmodel}) are used as priors
in our astrometric fits.   Although this model ignores the parallactic
perturbation of the astrometric signal, it still makes use of the
photometrically measured $\piE$.  The free parameters are: $\tnaught$,
$\tE$, $\uo$,  $\piE$, $\muSvec$, $\murelvec$ and $\Xsovec$, which is
the intrinsic source position on the plane of the sky $\left[\xpos,
  \ypos\right]$ at $t=t_0$, denoted by  East and North components.

The apparent position of the source in the sky plane is modeled as
\begin{equation}
\Xsvec\left(t\right) = \Xsovec + \muSvec\left[t-\tnaught\right] + \deltavec\left(t\right)
\end{equation}
where $\deltavec\left(t\right)$ is given by Equation \ref{eqn:astroshift2}.  
Both $\thetaE$ and $M$ are implicitly defined by each astrometric model via 
Equations \ref{eqn:tE} and \ref{eqn:mass} respectively.

The probability of different microlensing event models given our
astrometry is evaluated in the following Bayesian framework. 
According to Bayes theorem, the probability, $P$, of any microlensing
model $m$($\tnaught$, $\tE$, $\uo$, $\piE$,  $\muSvec$, $\murelvec$,
$\Xsovec$) given astrometric measurements $d \left(\xobs, \yobs \mid
  \tobs\right)$ is
\begin{equation}
P({m\mid d}) \propto P({d\mid m})P(m)
\label{eqn:Bayes}
\end{equation}
where $P(m)$ is the prior probability of $m$ and $P({d\mid m})$ is the
likelihood of the observed data given $m$. We adopt:
\begin{equation}
P({d\mid m}) = e^{-\chi^{2}/2}
\label{eqn:likelihood}
\end{equation}
where
\begin{equation}
\chi^{2} = \sum_{i}\left(\left[\frac{\xobsi-\xmodi}{\sigmaX}\right]^{2} + \left[\frac{\yobsi-\ymodi}{\sigmaY}\right]^{2} \right).
\end{equation}
Here, $\left[\xobsi, \yobsi\right]$ are the observed positions at
time $t^{obs}_i$ in the East and North directions,
$\left[\sigmaX, \sigmaY\right]$ are the corresponding astrometric
uncertainties on this observation, and $\left[\xmodi, \ymodi\right]$
are the predicted model positions.

The joint posterior distributions for $u_{0}$, $t_{0}$, $t_{E}$, and
$\piEvec$ derived from photometric modeling (\S \ref{sec:photmodel})
serve as priors for our astrometric microlensing models.   We adopt
uniform priors on all other model parameters. In particular, the
priors on the source and lens proper motions are uniform within the
range $[-40, 40]$ mas yr$^{-1}$.  While previous microlensing
studies have incorporated priors based on Galactic models, it is
unclear whether BHs have similar distance or velocity distributions as
luminous stellar populations.  The proper motions of the background
sources can be well-constrained by several years of astrometric
observations without relying on Galactic models.   

To compute parameter posterior distributions, we use a publicly
available Bayesian inference package, \texttt{MultiNest}
\citep{Feroz09, Feroz13}, which employs a nested sampling algorithm
\citep{Skilling06}.  \texttt{MultiNest} explores parameter space more
quickly and effectively than traditional MCMC algorithms when
parameters are strongly correlated and parameter space is multimodal.
For each sampling iteration, the Bayesian evidence is computed at a
fixed number of points \textit{N}, iteratively converging towards
smaller volumes of parameter space with the highest evidence.  The
routine terminates when the iterative increase in evidence falls below
some tolerance, \textit{E$_{\mathrm{tol}}$}.   We find that \textit{N}
= 1000, \textit{E$_{\mathrm{tol}}$}=0.3 produces adequate results.

\section{Results}

\label{sec:analysis}

\subsection{Proper motion fits:  No lensing}

\label{sec:propermotions}

\begin{figure}
\centering
\includegraphics[width=8cm]{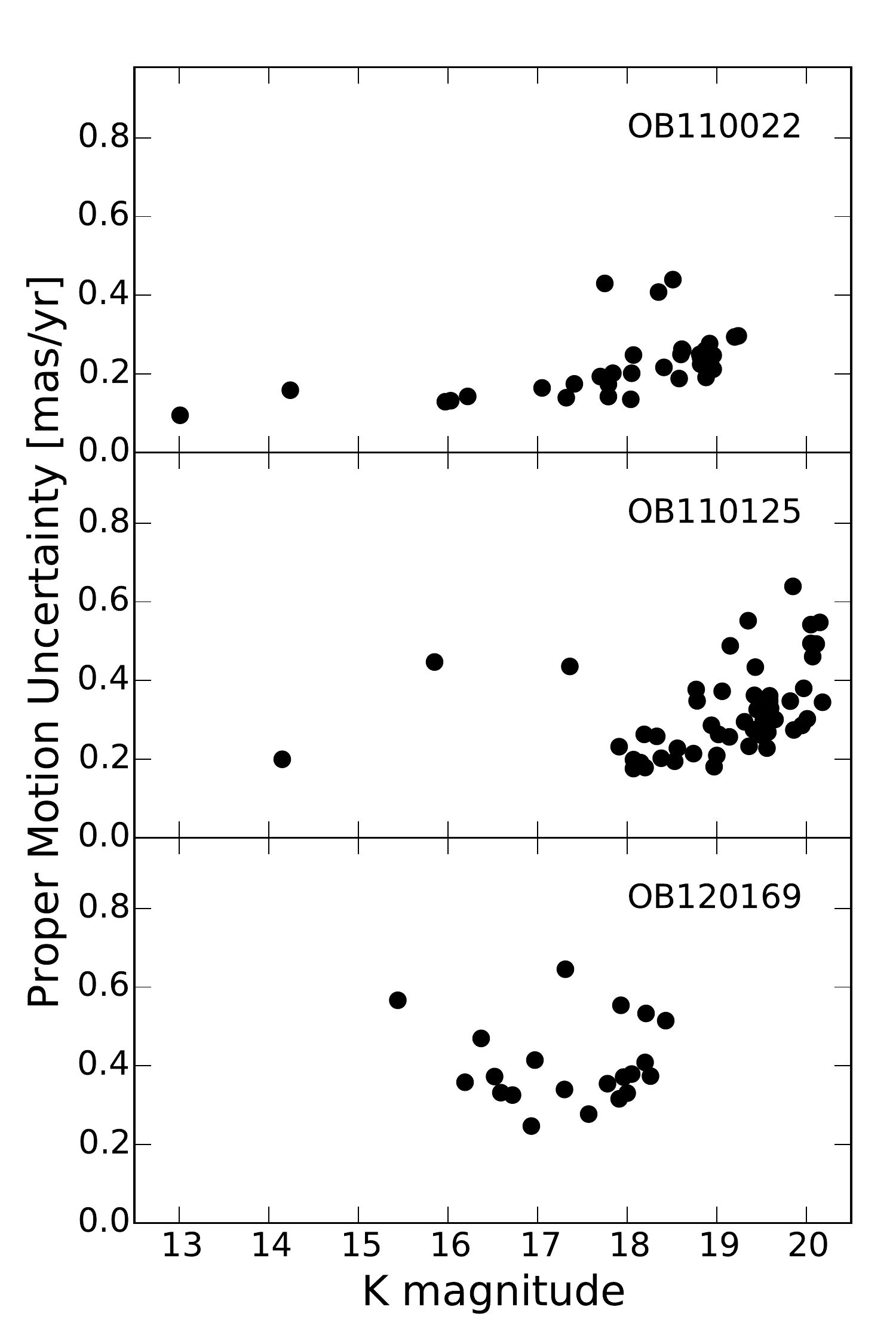}
\caption{Errors on the fitted proper motion plotted against source
  brightness for the OB110022 (top), OB110125 (middle) and OB120169
  (bottom) fields. Plotted errors are the average over X and Y.}
\label{fig:velerr}

\end{figure}

The resulting proper motions for the targets and all other stars
within 4\arcsec\; have precisions of $<$0.5 mas yr$^{-1}$ (Figure
\ref{fig:velerr}). The target with the highest quality data, OB110022,
has proper motion precisions of $<$0.2 mas yr$^{-1}$ for stars
brighter than $K<17$ mag. Given a reasonably stable atmosphere in
future experiments, proper motion errors of $<0.1$ mas yr$^{-1}$ are
obtainable.

%
%

\begin{figure*}
\centering
\includegraphics[width=18cm]{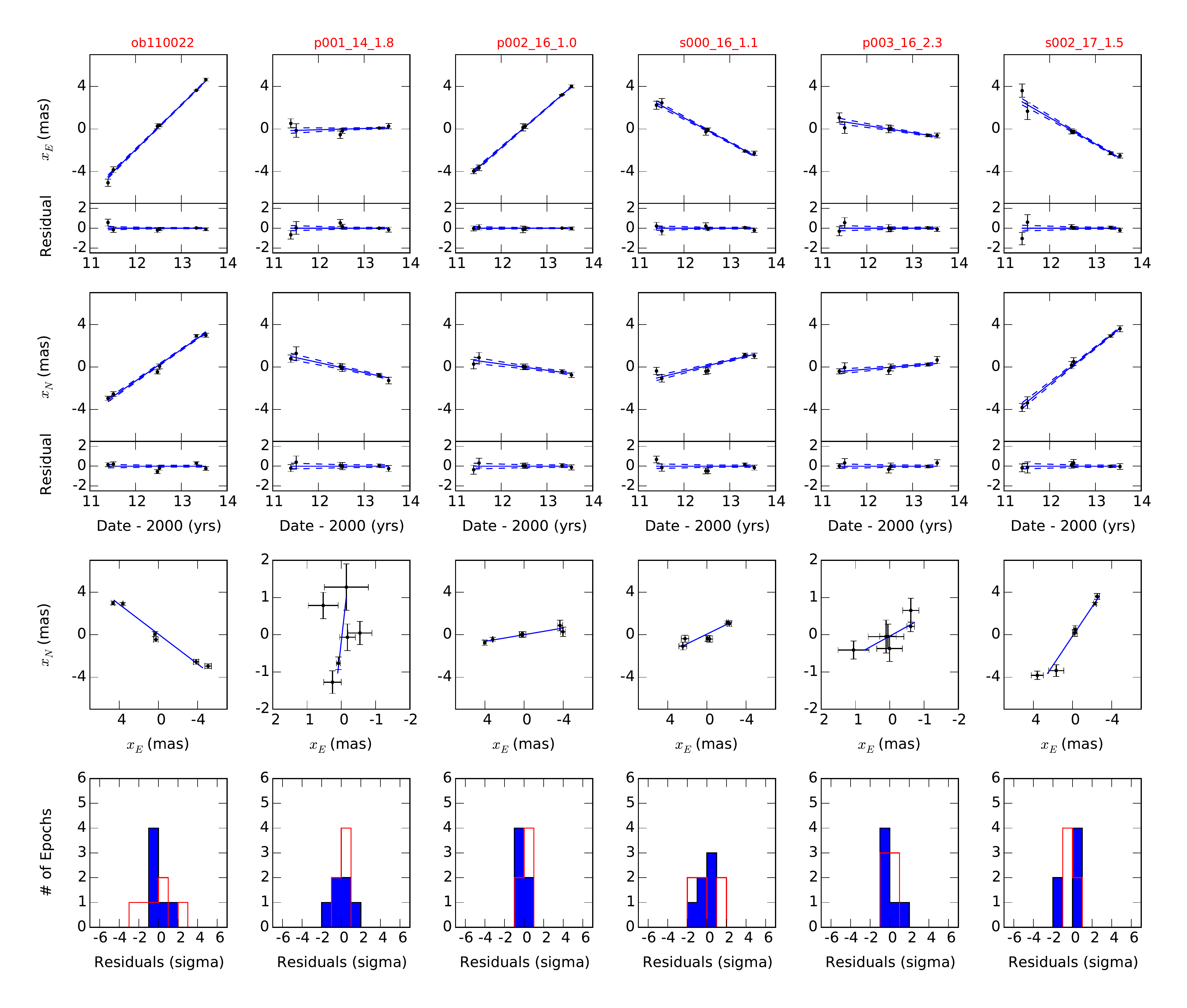}
\caption{Proper motion fits to OB110022 (leftmost column) and five comparison sources (one per column), which are both bright and near OB110022.   Sources names are given in red.  The three numbers in the names of the comparison sources correspond to a catalog number, K magnitude, and arcseconds from OB110022.   The rows are explained as follows:  \emph(Row 1) Linear proper motion fits to the East component, $\xpos$, of the astrometric time-series points (black).  The dashed blue line correspond to 1-sigma errors on the proper motion fit.  Corresponding residuals to the proper motion fit are shown.   \emph(Row 2) Same as Row 1 but for North component, $\ypos$, of astrometry time-series.   \emph(Row 3)  Linear fits to the proper motion in the sky plane.   \emph(Row 4) Histogram of residuals to the proper motion fits in $\xpos$ (red) and $\ypos$ (blue) in units of their 1-sigma error bars.}
\label{fig:ob110022}
\end{figure*}

\begin{figure*}
\centering
\includegraphics[width=18cm]{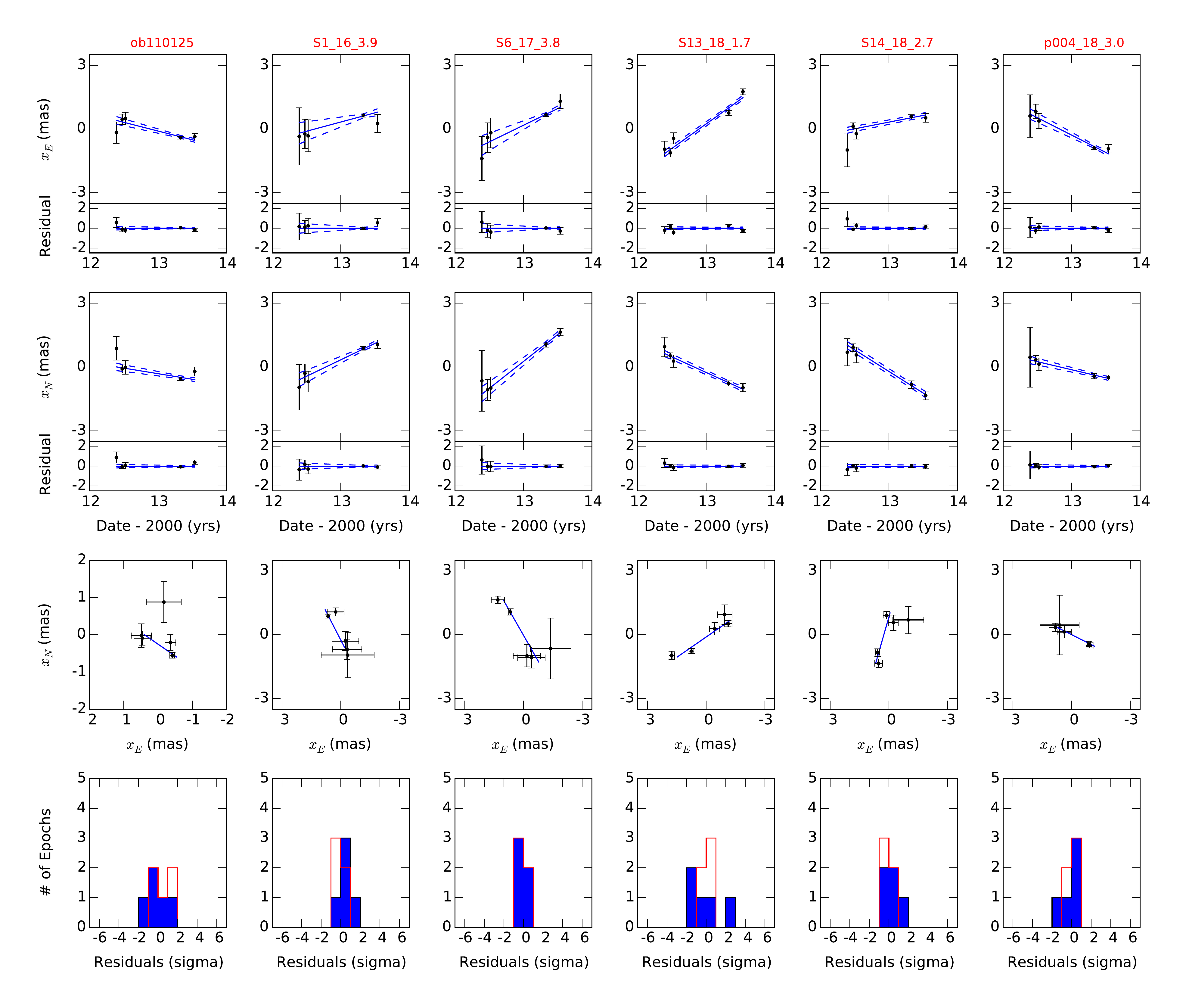}
\caption{Same as Figure \ref{fig:ob110022} but for the OB110125 field.}
\label{fig:ob110125}
\end{figure*}

\begin{figure*}
\centering
\includegraphics[width=18cm]{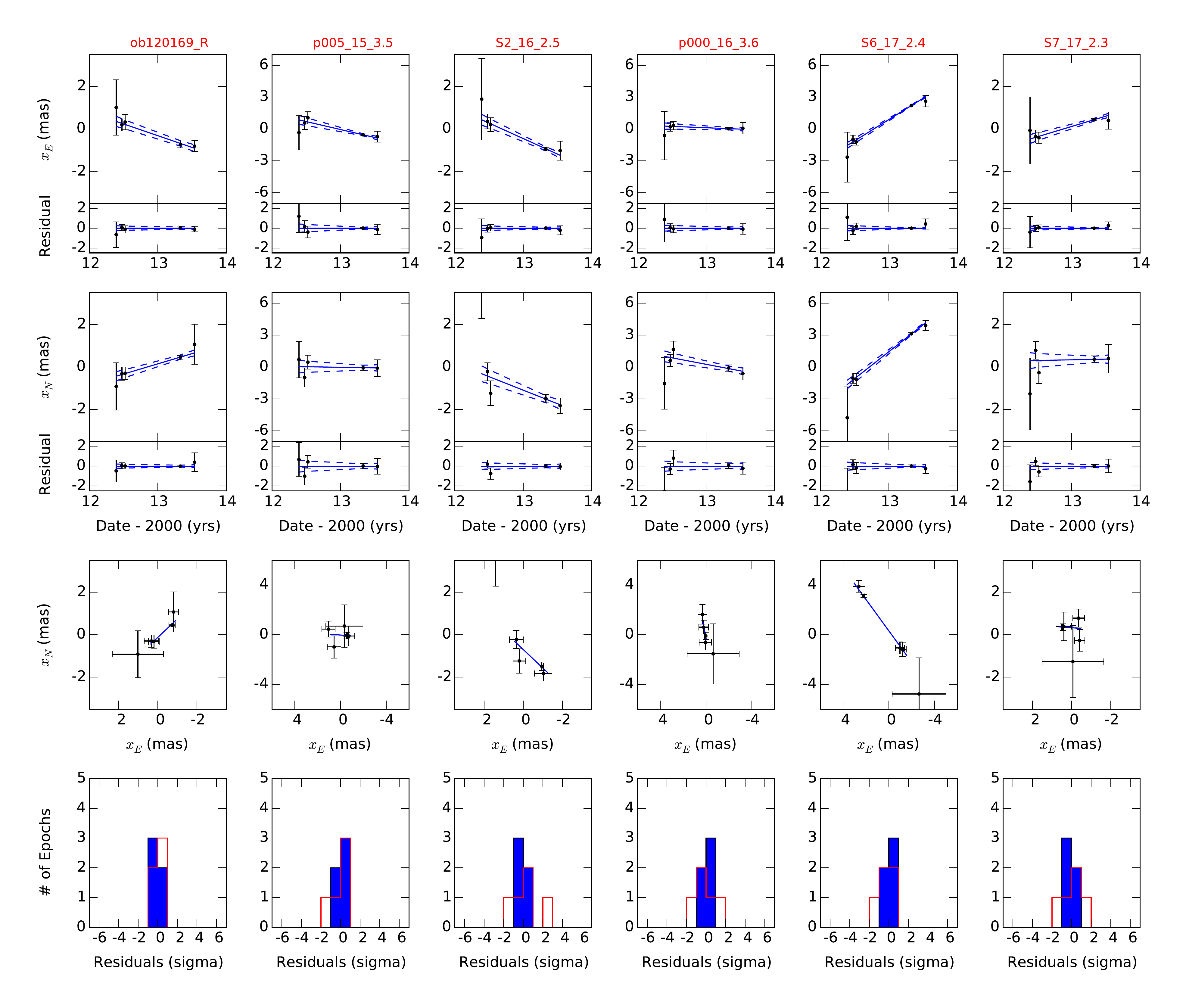}
\caption{Same as Figure \ref{fig:ob110022} but for the OB120169 field.}
\label{fig:ob120169}
\end{figure*}

The best-fit proper motions for the three targets of interest
and the associated errors are shown in Table \ref{tb:pm}.
Figures \ref{fig:ob110022}-\ref{fig:ob120169} show the proper motion
fits for the targets and five of the brightest stars closest to each
of them, which serve as comparison samples.  The corresponding
residuals are also shown.  There are clearly some residual systematics
affecting each target uniquely, but the consistency of the entire
ensemble of residuals with the measured Gaussian errors suggests that
we cannot do much better given the poor atmospheric conditions under
which much of the data was obtained.  The relatively comparable
frequency of outliers in the comparison sample relative to the target
of interest suggests that points deviating far from the proper motion
fit cannot be interpreted as a lens induced signature without proper
astrometric modeling and comparison with
a reference sample.

\begin{figure*}
\includegraphics[scale=0.38]{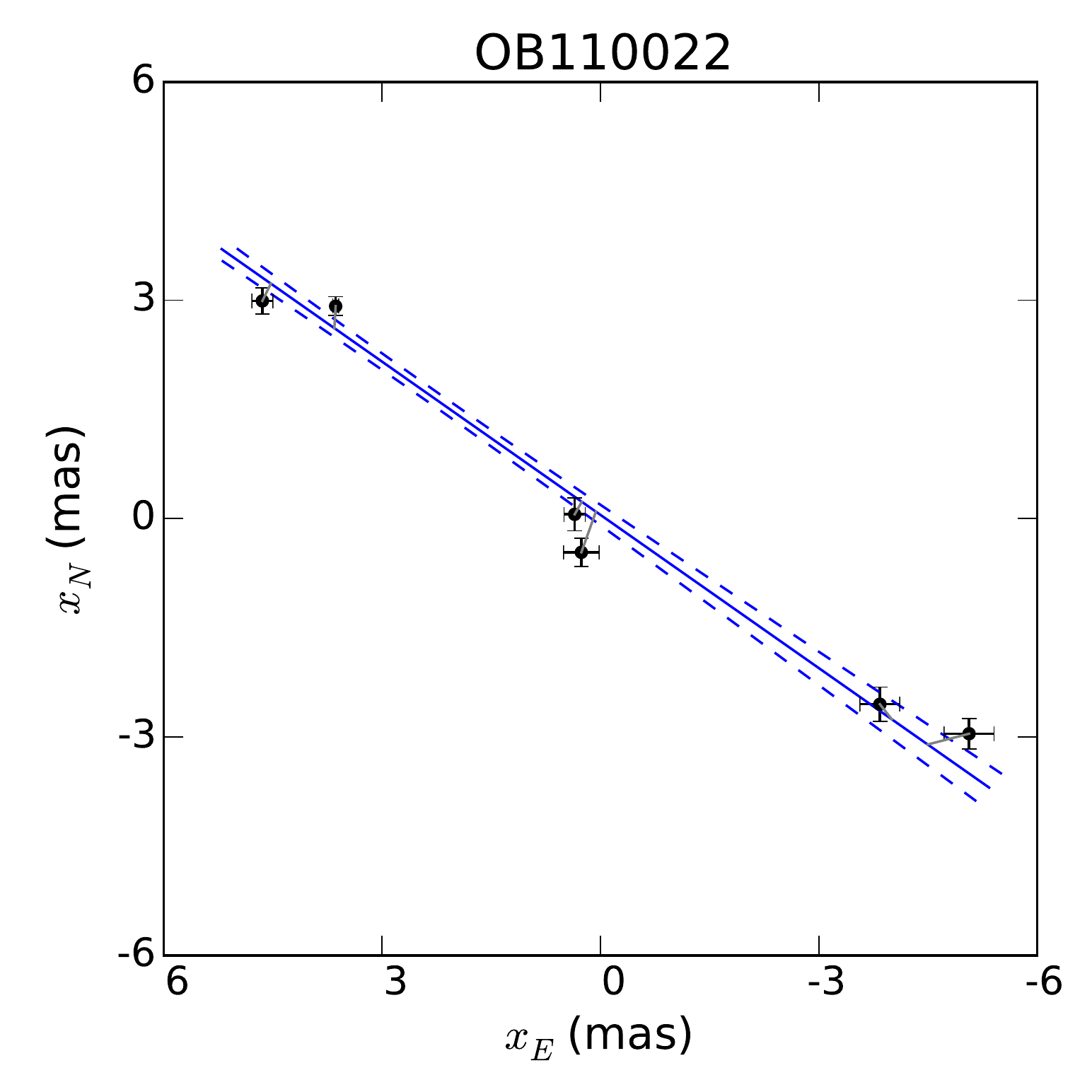}
\includegraphics[scale=0.38]{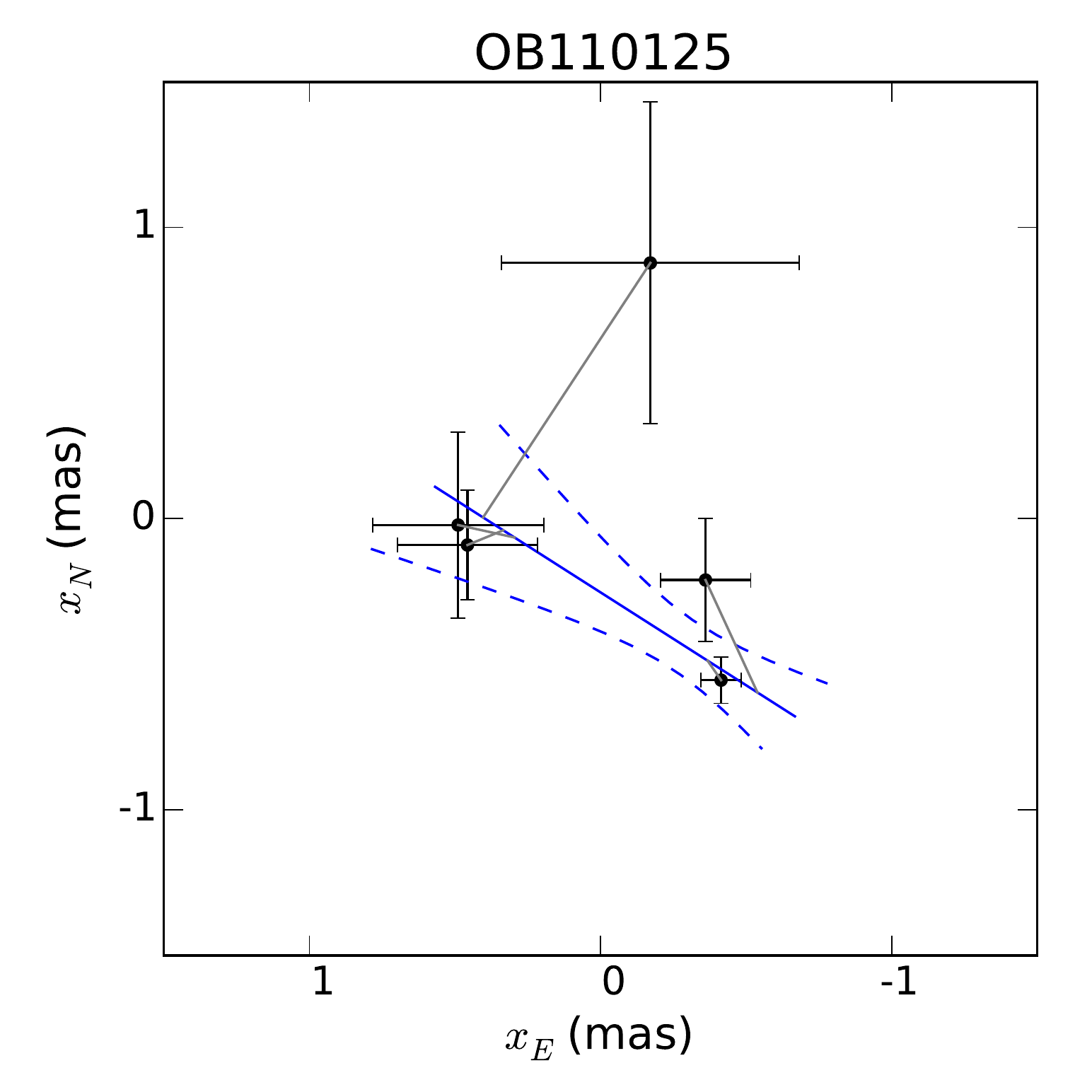}
\includegraphics[scale=0.38]{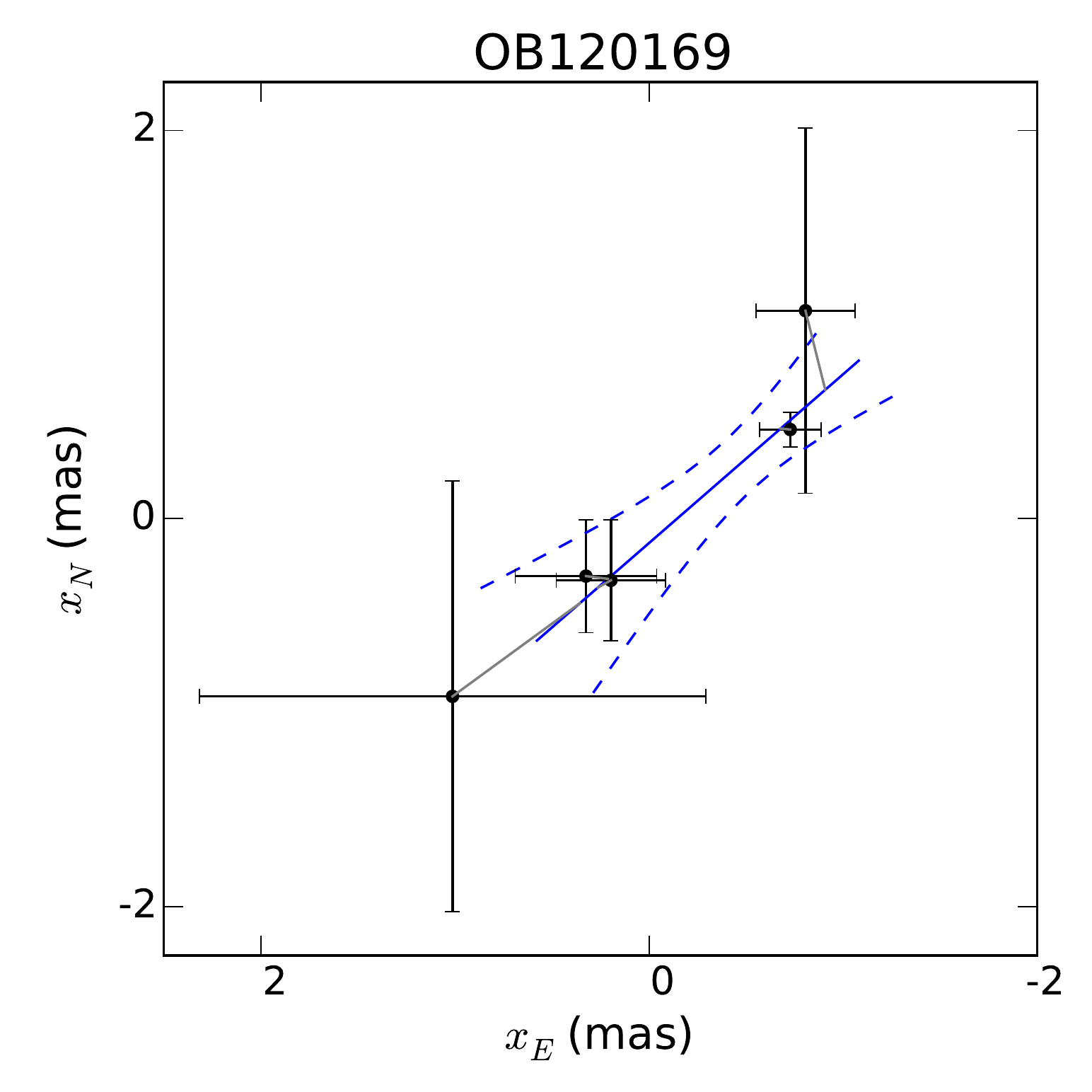}
\caption{
A zoomed-in view of the proper motion fits for OB110022 ({\em left}),
OB110125 ({\em middle}), and OB120169 ({\em right}). Observed positions on the sky plane are shown as {\em black points}. The best-fit proper motion and 1$\sigma$ uncertainties are shown in {\em blue}.
}
\label{fig:prop_mot_zoom}
\end{figure*}

%
The three targets of interest are all consistent with linear motion
(Figure \ref{fig:prop_mot_zoom}).
The $\chi^2$ of the target's proper motion fit is not a significant
outlier when compared to all other stars in the field with K$<$22 mag
and detected in all epochs.
Specifically, OB110022 has a $\chi^2=22.2$ (for 8 degrees of freedom
(DOF)) and 21\% of the comparison sample have higher $\chi^2$ values,
OB110125 has a $\chi^2=27.6$ (for 6 DOF) and 16\% of the comparison
sample have higher $\chi^2$ values, and OB120169 has a
$\chi^2=1.1$ (for 6 DOF) and 86\% of the comparison sample have higher
$\chi^2$ values.
The distributions of $\chi^2$ values from the proper motion fits for
the complete sample of reference stars
in the fields around all three targets are plotted
in Figure \ref{fig:hist_chi2_pa} (Appendix \ref{sec:align_order}). 

While we do not detect significant astrometric microlensing in any of
the three targets, the strong limits on any non-linearity in the case
of OB110022 allow us to constrain the properties of the source-lens
system as shown in \S\ref{sec:ob110125_results}.
We note that although the $\chi^2$ distribution for OB110022
is skewed to higher values than expected for the adopted O=1
transformation, the larger astrometric error bars produced in fits 
using O=2 produced a negligable change in the final lens-mass posteriors.

%
\begin{deluxetable}{lrrrc}
\tabletypesize{\footnotesize}
\tablecaption{Proper Motions}
\tablecolumns{3}
\tablewidth{0pt}
\tablehead{
Source & $\mu_{\x}$ & $\mu_{\y}$  \\ 
& [mas yr$^{-1}$] & [mas yr$^{-1}$] & $\chi_{vel}^2$ & DOF \\ 
}
\startdata
OB110022   &  4.22 $\pm$  0.09 &  2.96 $\pm$  0.09 &  22.2 & 8  \\
OB110125   & -0.82 $\pm$  0.20 & -0.53 $\pm$  0.19 &  10.4 & 6  \\
OB120169   & -1.11 $\pm$  0.28 &  0.96 $\pm$  0.27 &   1.1 & 6 
\enddata
\label{tb:pm}
\end{deluxetable}

%
%

%
%

%
%

%
%

\subsection{Lens Masses}

\subsubsection{OB120169}

\begin{figure}
\centering
\includegraphics[width=0.9\columnwidth]{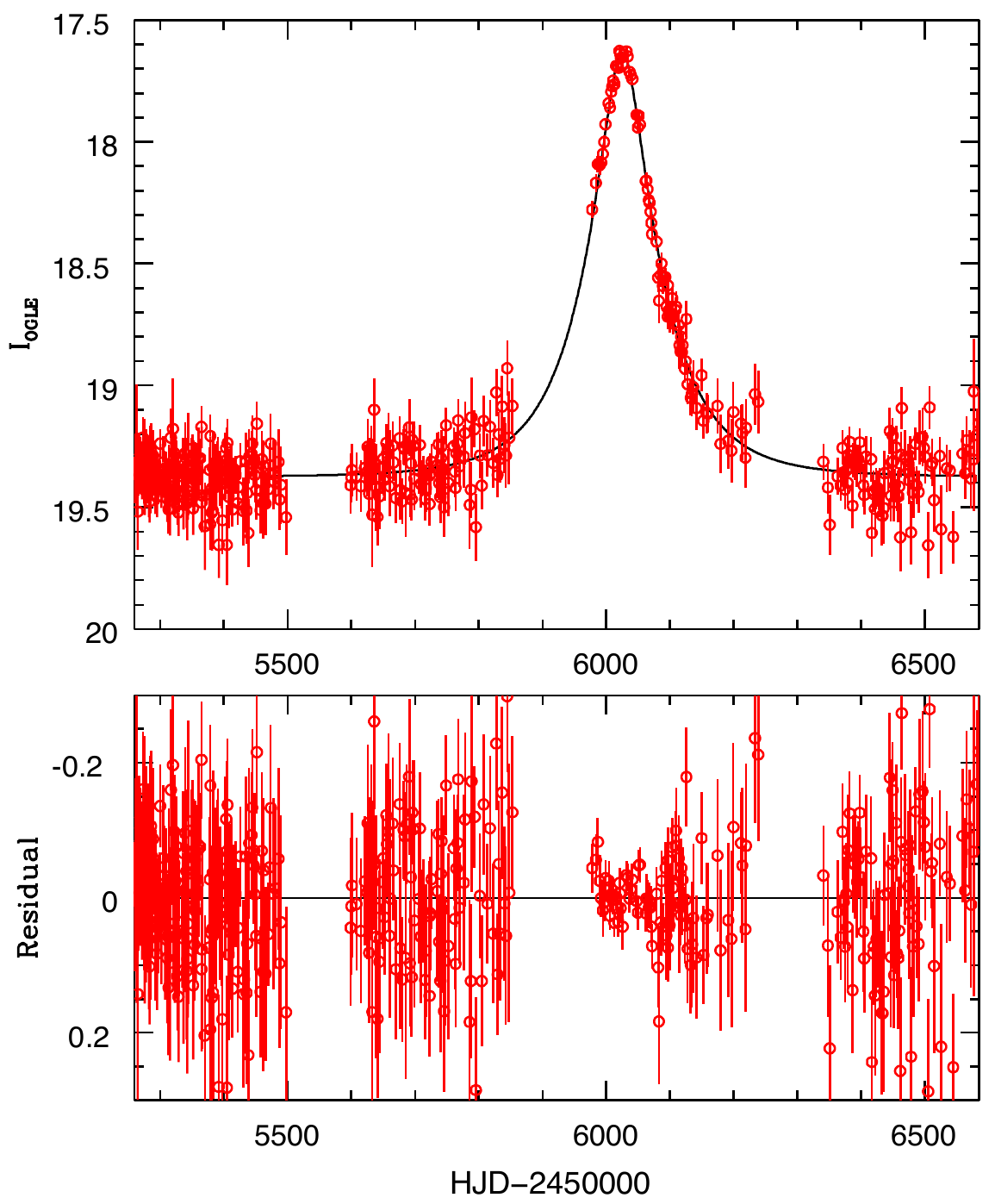}
\caption{\emph{Top:} Light curve of OB120169.  Red points indicate
  OGLE measurements.  The black line represents the best-fitting
  point-source-point-lens model, which includes microlensing
  parallax. \emph{Bottom}:  Residuals to best-fitting model.}
\label{fig:OB120169lc}
\end{figure}

Figure \ref{fig:OB120169lc} shows the OGLE $I$-band light curve as
well as the best-fitting photometric microlensing model, which has a
single, isolated lens.  Table \ref{tb:parsOB120169} lists the medians
and 68\% confidence intervals of the posterior distributions for each
model parameter derived from photometry and from astrometric models
informed by photometric priors.  We include both the $u_0^+$ and
$u_0^-$ solutions because neither is strongly preferred in the
fit. 
%
While we utilize the Bayesian evidence in our fitting procedure, it is
useful to compare the resulting $\chi^2$ of the best-fit astrometric
lensing solution ($\chi^2=1.13$ for 1 DOF) to that of the linear
proper motion fit ($\chi^2=1.10$ for 6 DOF). 
In this
case, the lensing model did not yield a significant improvement in the
$\chi^2$ value over the linear proper motion model.

%
%
%
\begin{figure*}
\centering
\includegraphics[width=10cm]{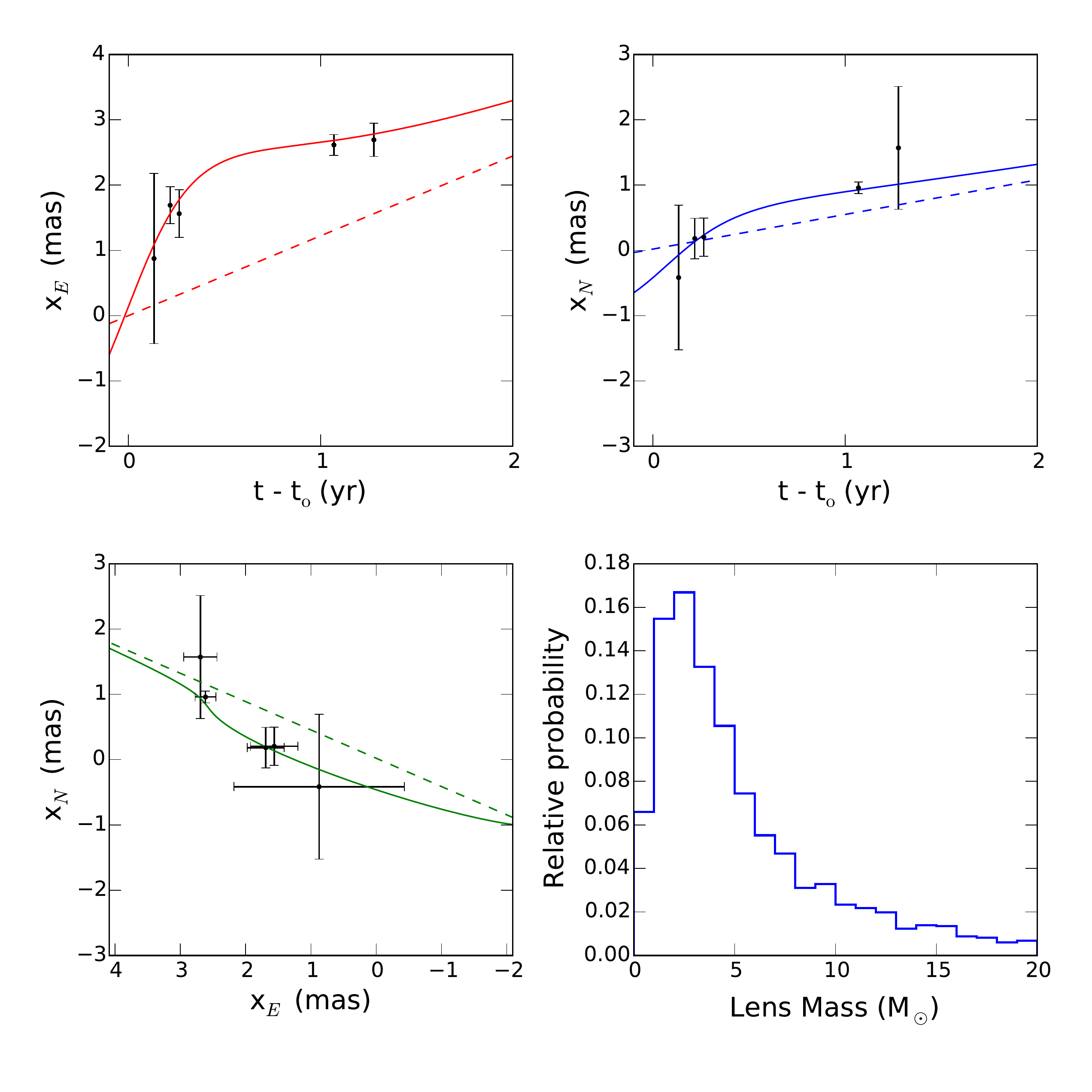}
\label{fig:bestfit120169-}
\caption{\emph{Left}:  \emph{Top}:  Fitted astrometric data (black
  points) for OB120169, with the best-fitting astrometric model
  over-plotted in $\xpos$ (red solid line) and $\ypos$ (blue solid
  line) versus time after minimum source-lens separation.  Dashed
  lines indicate unlensed source motions.  The joint posterior
  probability distributions for $u_{0} < 0$, $t_{0}$, $t_{E}$,
  $\pi_{E,N}$, and $\pi_{E}$ from light-curve fitting are adopted as
  priors for astrometric model fitting.   A single single
  point-source-point-lens astrometric model is used that ignores the
  effects of source-lens relative parallax.  \emph{Bottom Left}:  Same
  as top, but shown in the sky plane.   \emph{Bottom Right}:  Lens
  mass posterior probability distribution.}
\label{fig:model120169-}
\end{figure*}

%
%
%
\begin{figure*}
\centering
\includegraphics[width=10cm]{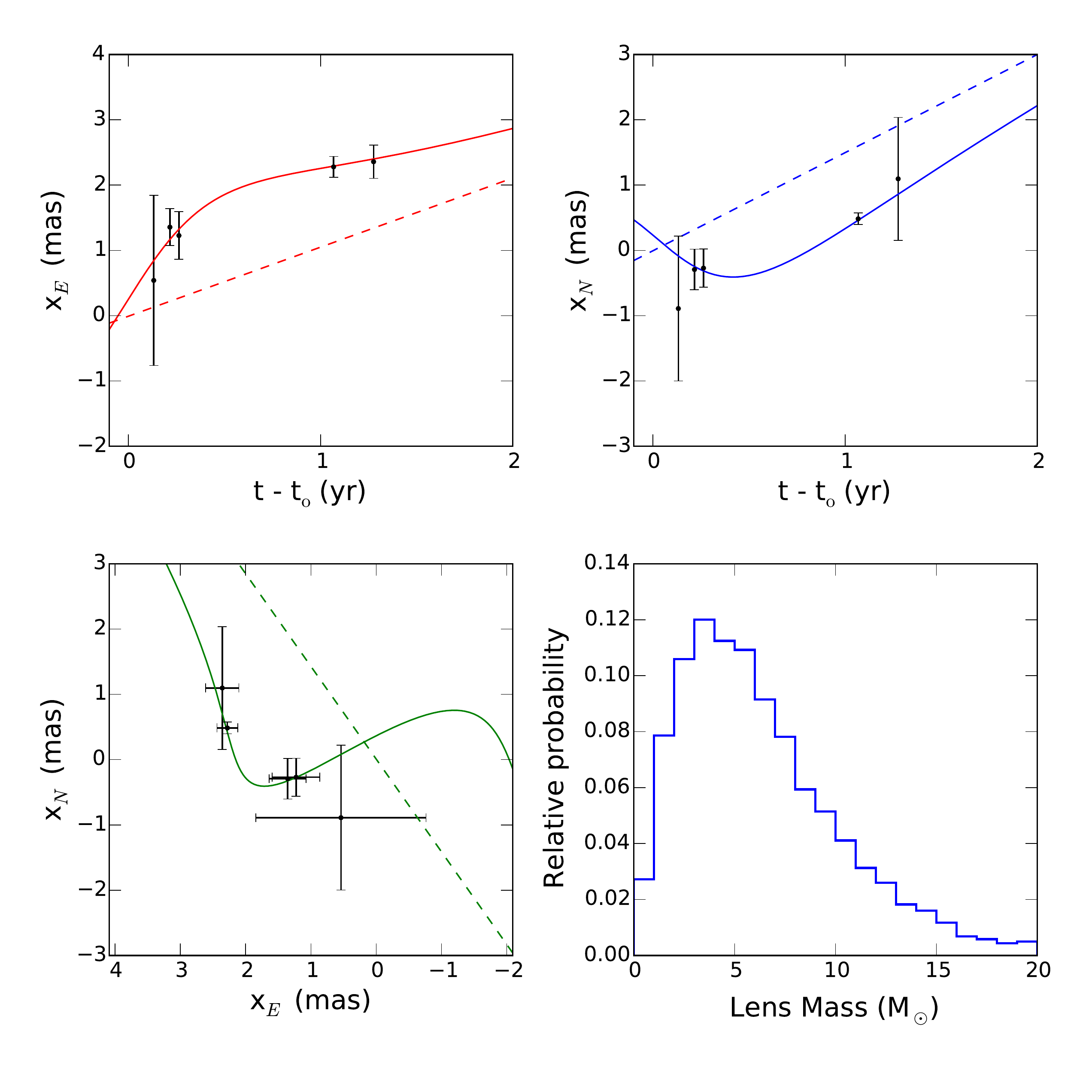}
\label{fig:bestfit120169+}
\caption{Same as Figure \ref{fig:model120169-} but adopting the joint
  posterior probability distributions for $u_{0} > 0$, $t_{0}$,
  $t_{E}$, and $\pi_{E}$ from our light-curve fitting as priors for
  our astrometric model fitting.}
\label{fig:model120169+}
\end{figure*}

Figures \ref{fig:model120169-} and \ref{fig:model120169+} display the
best fitting astrometric microlensing models and lens mass constraints
corresponding to the $u_0^+$ and $u_0^-$ solutions respectively.
The lens mass posterior associated with the $\uominus$ solution has a
%
%
median of 5.7 M$\subsun$, but wide 1-sigma and 3-sigma confidence
intervals of 4.0--7.7 and 0.4--39.8 M$\subsun$ respectively.  The
$\uoplus$ solution has a median of 4.0 M$\subsun$ and similarly wide
1-sigma and 3-sigma confidence intervals of 2.6--6.2 and 0.2--38.8
M$\subsun$ respectively. The best-fitting models for $\uominus$ and
$\uoplus$ solutions are shown in Figures \ref{fig:model120169-} and
\ref{fig:model120169+}.  Both solutions predict that the y-component
of the source apparent motion has started to turnover and approach
linear, unlensed motion.  Future observations are essential to
determine if this turnover continues and will place a more conclusive
constraint on the lens mass, possibly verifying a black hole.

%
%
\begin{deluxetable*}{lrrrr}[h]
\centering
\tabletypesize{\footnotesize}
\tablecaption{Microlensing Event Parameters - OB120169}
\centering
\tablecolumns{5}
\tablehead{
& \multicolumn{2}{c}{$\uominus$ Solution} & \multicolumn{2}{c}{$\uoplus$ Solution} \\
Parameter & \multicolumn{1}{c}{Photometry} & \multicolumn{1}{c}{Astrometry} & \multicolumn{1}{c}{Photometry} & \multicolumn{1}{c}{Astrometry}
}
\startdata
$t_0$ (HJD - 2450000)                         &      6026.03$^{+        0.43}_{-        0.40}$ &      6026.04$^{+        0.28}_{-        0.28}$ &      6026.25$^{+        0.43}_{-        0.38}$ &      6026.27$^{+        0.29}_{-        0.26}$ \\ [0.2cm]
$u_0$                                         &       -0.222$^{+       0.033}_{-       0.031}$ &       -0.229$^{+       0.024}_{-       0.025}$ &        0.166$^{+       0.044}_{-       0.070}$ &        0.165$^{+       0.026}_{-       0.042}$ \\ [0.2cm]
$t_E$ (days)                                  &          135$^{+          21}_{-          36}$ &          121$^{+           9}_{-          15}$ &          156$^{+          32}_{-          42}$ &          157$^{+          23}_{-          21}$ \\ [0.2cm]
$\pi_{E,E}$                                   &       -0.058$^{+       0.017}_{-       0.018}$ &       -0.057$^{+       0.014}_{-       0.014}$ &       -0.031$^{+       0.018}_{-       0.016}$ &       -0.030$^{+       0.012}_{-       0.011}$ \\ [0.2cm]
$\pi_{E,N}$                                   &         0.11$^{+        0.30}_{-        0.12}$ &        -0.11$^{+        0.19}_{-        0.23}$ &         0.14$^{+        0.11}_{-        0.05}$ &         0.14$^{+        0.06}_{-        0.04}$ \\ [0.2cm]
$\mu_{s,E}~(\mathrm{mas~yr^{-1})}$            &          ---                                   &        -1.15$^{+        0.48}_{-        0.53}$ &          ---                                   &        -0.96$^{+        0.53}_{-        0.51}$ \\ [0.2cm]
$\mu_{s,N}~(\mathrm{mas~yr^{-1})}$            &          ---                                   &         0.85$^{+        0.51}_{-        0.52}$ &          ---                                   &         1.05$^{+        0.64}_{-        0.63}$ \\ [0.2cm]
$\mu_{\mathrm{rel},E}~(\mathrm{mas~yr^{-1})}$ &          ---                                   &         -3.8$^{+        14.7}_{-        14.5}$ &          ---                                   &         -4.2$^{+        10.8}_{-        11.5}$ \\ [0.2cm]
$\mu_{\mathrm{rel},N}~(\mathrm{mas~yr^{-1})}$ &          ---                                   &          1.2$^{+        14.9}_{-        15.0}$ &          ---                                   &          1.5$^{+        13.4}_{-        13.8}$ \\ [0.2cm]
$\theta_E$ (mas)                              &          ---                                   &          5.9$^{+         3.0}_{-         3.9}$ &          ---                                   &          6.3$^{+         3.3}_{-         4.4}$ \\ [0.2cm]
Mass ($M_{\odot}$)                            &          ---                                   &          4.0$^{+         2.3}_{-         6.4}$ &          ---                                   &          5.6$^{+         3.1}_{-         5.1}$ \\ [0.2cm]
$I_{\mathrm{OGLE}}$                           &       19.266$^{+       0.181}_{-       0.195}$ &          ---                                   &       19.641$^{+       0.456}_{-       0.386}$ &          ---                                   \\ [0.2cm]
$f_{b}/f_{s}$                                 &        -0.10$^{+        0.14}_{-        0.17}$ &          ---                                   &         0.27$^{+        0.43}_{-        0.54}$ &          ---                                   \\ [0.2cm]
$\chi^{2}$                                    &        428.5                                   &          1.1                                   &        424.7                                   &          1.3                                   \\ [0.2cm]
N$_{\mathrm{dof}}$                            &          433                                   &            1                                   &          433                                   &            1                                   \\ [0.2cm]
\enddata
\tablenotetext{}{}
\label{tb:parsOB120169}
\end{deluxetable*}

\subsubsection{OB110022}
\label{sec:ob110122_results}

\begin{figure}
\centering
\includegraphics[width=0.9\columnwidth]{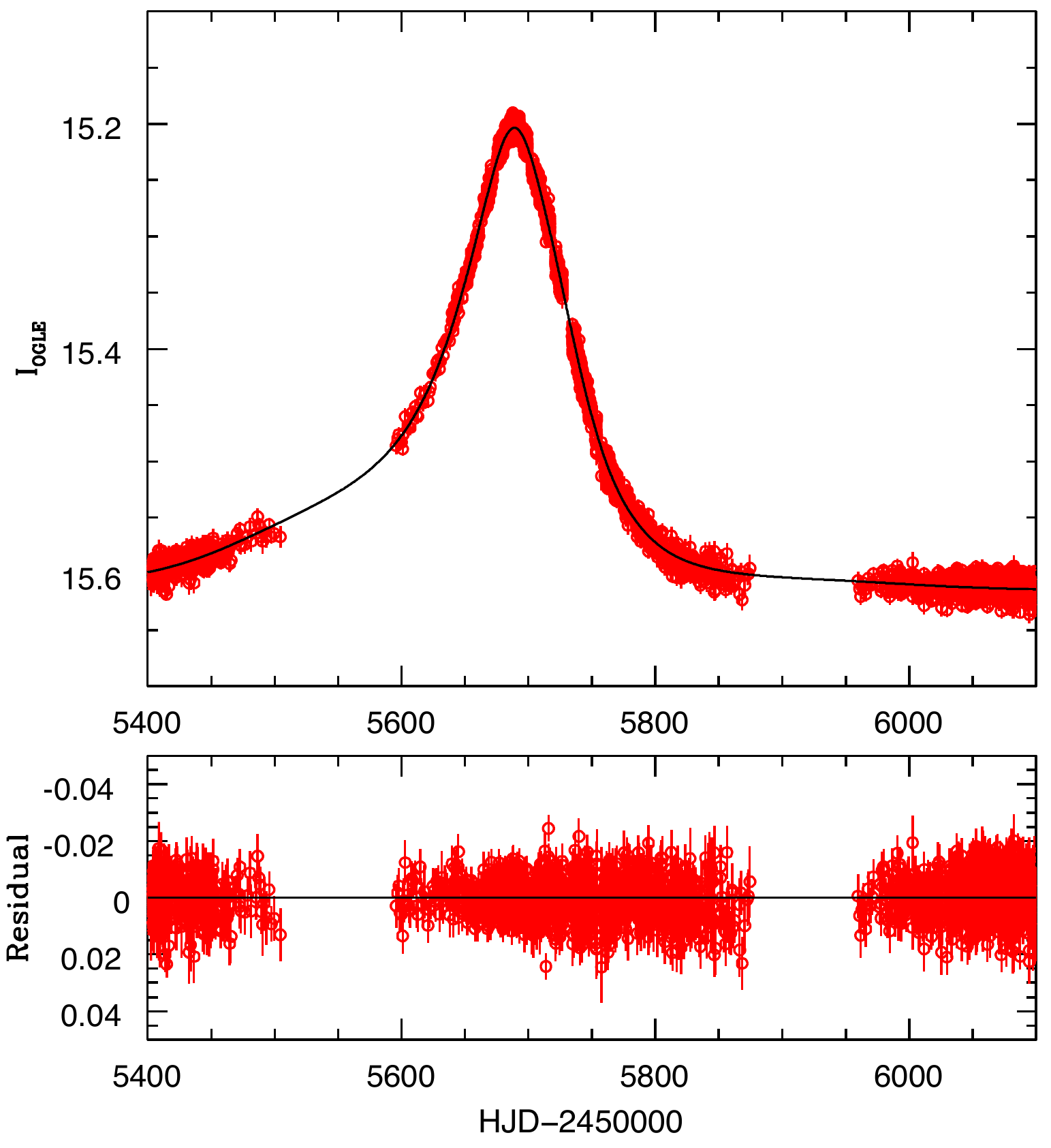}
\caption{ \emph{Top}: Light curve of OB110022.  Red points indicate
  OGLE measurements.  The black line represents the best-fitting
  binary lens model, which includes microlensing parallax and binary
  orbital motion.  \emph{Bottom}:  Residuals to best-fitting model.}
\label{fig:OB110022lc}
\end{figure}

Figure \ref{fig:OB110022lc} shows the OGLE $I$-band light curve as
well as the best-fitting photometric microlensing model, which has a
non-static, binary lens.  Table \ref{tb:parsOB110022} lists the
posterior medians and 68\% confidence intervals for each OB110022
event parameter.  We separately list measurements derived from
photometry only, and from astrometric modeling informed by photometric
priors.  The blended flux $\fb$ is comparable to the source flux
$\fsrc$. However, we note that $\fb$ increased substantially when
adopting the non-static solution, which may indicate that more
detailed modeling is required.   The median of the posterior for
binary separation $s$ = 0.42 suggesting that the binary separation is
small compared to the Einstein radius.
The light curve is relatively smooth indicating that there are no
caustic approaches or crossings that would lead to additional,
binary-induced non-linearity in the apparent source trajectory
\citep{Sajadian14}.
Additionally, no such non-linearity is detected in the astrometry  
and the highly linear proper motion measured for OB110022 is therefore
consistent with its smooth light curve.

\begin{figure*}
\centering
\includegraphics[width=10cm]{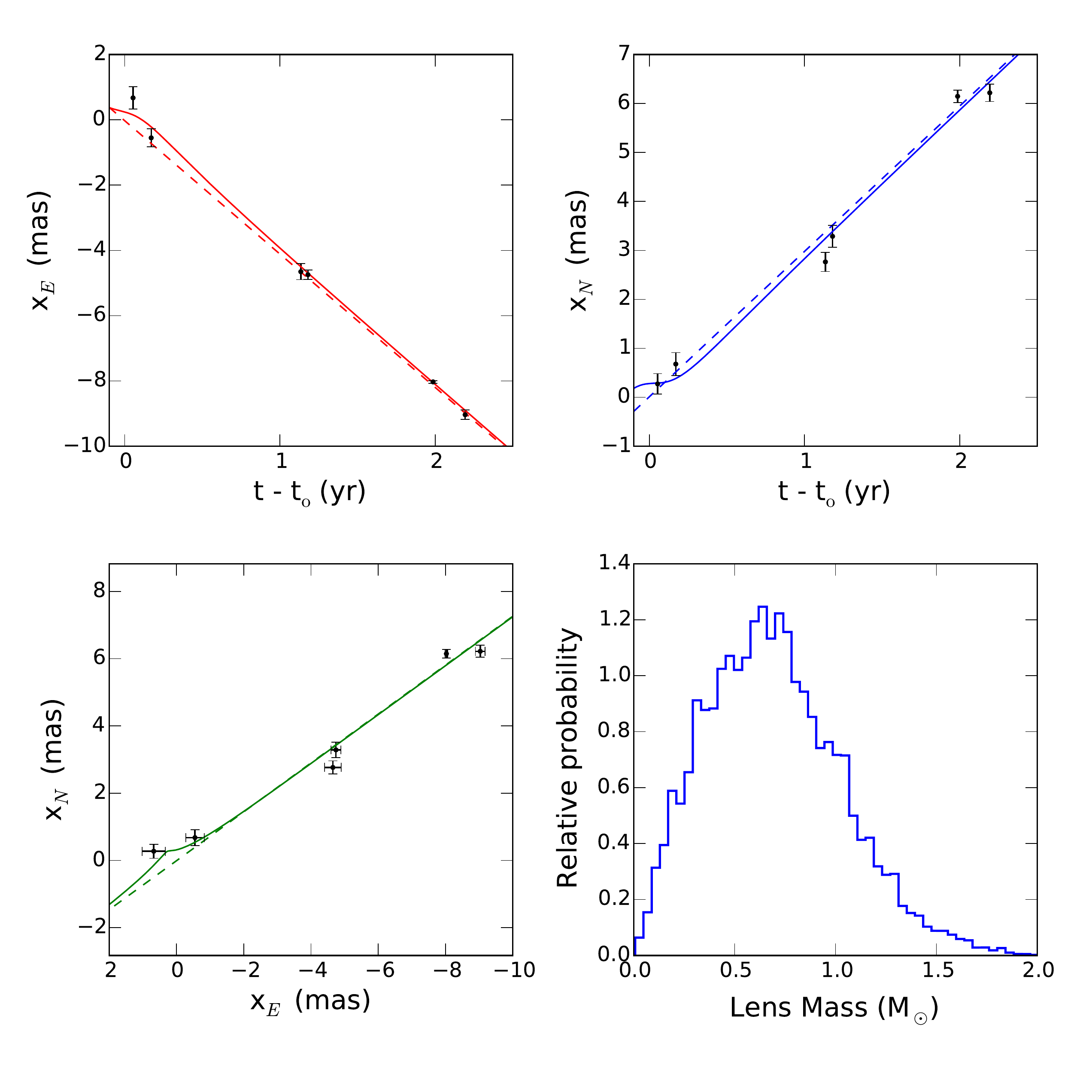}
\label{fig:bestfit110022}
\caption{\emph{Top}:  Fitted astrometric data (black points) for
  OB110022, with the best-fitting astrometric model over-plotted in
  $\xpos$ (red solid line) and $\ypos$ (blue solid line) versus time
  after minimum source-lens separation.  Dashed lines indicate
  unlensed source motions.  The joint posterior probability
  distributions for $u_{0} $, $t_{0}$, $t_{E}$, $\pi_{E,N}$, and
  $\pi_{E}$ from light-curve fitting are adopted as priors for
  astrometric model fitting.  A single single point-source-point-lens
  astrometric model is used that ignores the effects of source-lens
  relative parallax.  \emph{Bottom Left}:  Same as top, but shown in
  the sky plane.   \emph{Bottom Right}:  Lens mass posterior
  probability distribution.}  
\label{fig:model110022}
\end{figure*}

%
No significant astrometric lensing signal is detected. 
The best-fitting lensing model for the astrometry gives a $\chi^2 =
21.2$ (3 DOF), 
which is not a significant improvement over
the linear proper motion model ($\chi^2 = 22.2$ for 8 DOF).
Given that the lensing event was well sampled during the period of
maximum astrometric shift (Eq.~\ref{eqn:astroshift1}), 
the degree of linearity allows us to put constraints on the lens mass.
Figure \ref{fig:model110022} displays the posterior probability
distribution for the OB110022 lens mass and the best-fitting
astrometric model.   Most notably, the posterior corresponding to the
%
%
total lens mass $M$ has a median of 0.67 M$\subsun$ with 1-$\sigma$
and 3-$\sigma$ confidence intervals of 0.51--0.83 M$\subsun$ and
0.05--1.79 M$\subsun$, respectively. The highest masses correspond
to large relative proper motions between the source and lens
($\murel > 20$ mas yr$^{-1}$ for M$>$1 \msun). 
Combining $M$ and $q$, the
individual masses of the binary lens objects are $\sim$0.5 M$\subsun$
and 0.1 M$\subsun$, consistent with some combination of K- or M-dwarfs
and white dwarfs.  This lens is extremely unlikely to contain a black
hole.  Constraint of the lens objects to such low masses is not
surprising given the high degree of linearity of the measured source
motion, combined with relatively low astrometric error and multi-year
time baseline.

%
%
%

%
%
\begin{deluxetable}{lrr}[h]
\tabletypesize{\footnotesize}
\centering
\tablecaption{Microlensing Event Parameters - OB110022}
\tablecolumns{3}
\tablewidth{0.9\columnwidth}
\tablehead{
\colhead{Parameter} &
\colhead{Photometry} &
\colhead{Astrometry}
}
\startdata
$t_0$ (HJD - 2450000)                         &      5687.91$^{+        0.26}_{-        0.27}$ &      5687.41$^{+        0.25}_{-        0.26}$ \\ [0.2cm]
$u_0$                                         &        0.573$^{+       0.019}_{-       0.020}$ &        0.574$^{+       0.018}_{-       0.019}$ \\ [0.2cm]
$t_E$ (days)                                  &         61.4$^{+         1.0}_{-         1.0}$ &         61.4$^{+         0.9}_{-         0.9}$ \\ [0.2cm]
$\pi_{E,E}$                                   &       -0.393$^{+       0.013}_{-       0.012}$ &       -0.393$^{+       0.013}_{-       0.012}$ \\ [0.2cm]
$\pi_{E,N}$                                   &       -0.071$^{+       0.013}_{-       0.014}$ &       -0.071$^{+       0.014}_{-       0.014}$ \\ [0.2cm]
$\mu_{s,E}~(\mathrm{mas~yr^{-1})}$            &          ---                                   &         4.06$^{+        0.16}_{-        0.16}$ \\ [0.2cm]
$\mu_{s,N}~(\mathrm{mas~yr^{-1})}$            &          ---                                   &         3.02$^{+        0.20}_{-        0.20}$ \\ [0.2cm]
$\mu_{\mathrm{rel},E}~(\mathrm{mas~yr^{-1})}$ &          ---                                   &         -7.3$^{+         8.7}_{-         8.8}$ \\ [0.2cm]
$\mu_{\mathrm{rel},N}~(\mathrm{mas~yr^{-1})}$ &          ---                                   &         -5.5$^{+         8.0}_{-         7.6}$ \\ [0.2cm]
$\theta_E$ (mas)                              &          ---                                   &         2.19$^{+        1.06}_{-        1.17}$ \\ [0.2cm]
Mass ($M_{\odot}$)                            &          ---                                   &         0.67$^{+        0.32}_{-        0.37}$ \\ [0.2cm]
s                                             &        0.423$^{+       0.006}_{-       0.006}$ &          ---                                   \\ [0.2cm]
q                                             &        0.191$^{+       0.009}_{-       0.010}$ &          ---                                   \\ [0.2cm]
$\alpha$                                      &        4.794$^{+       0.014}_{-       0.015}$ &          ---                                   \\ [0.2cm]
$\omega$                                      &        0.062$^{+       0.006}_{-       0.007}$ &          ---                                   \\ [0.2cm]
$\dot{s}/s$                                   &        0.468$^{+       0.015}_{-       0.020}$ &          ---                                   \\ [0.2cm]
$I_{\mathrm{OGLE}}$                           &       16.264$^{+       0.057}_{-       0.054}$ &          ---                                   \\ [0.2cm]
$f_{b}/f_{s}$                                 &         0.82$^{+        0.73}_{-        0.91}$ &          ---                                   \\ [0.2cm]
$\chi^{2}$                                    &         9235                                   &         20.8                                   \\ [0.2cm]
N$_{\mathrm{dof}}$                            &         8253                                   &            3                                   \\ [0.2cm]
\enddata
\tablenotetext{}{}
\label{tb:parsOB110022}
\end{deluxetable}

\subsubsection{OB110125}
\label{sec:ob110125_results}

\begin{figure}
\centering
\includegraphics[width=0.9\columnwidth]{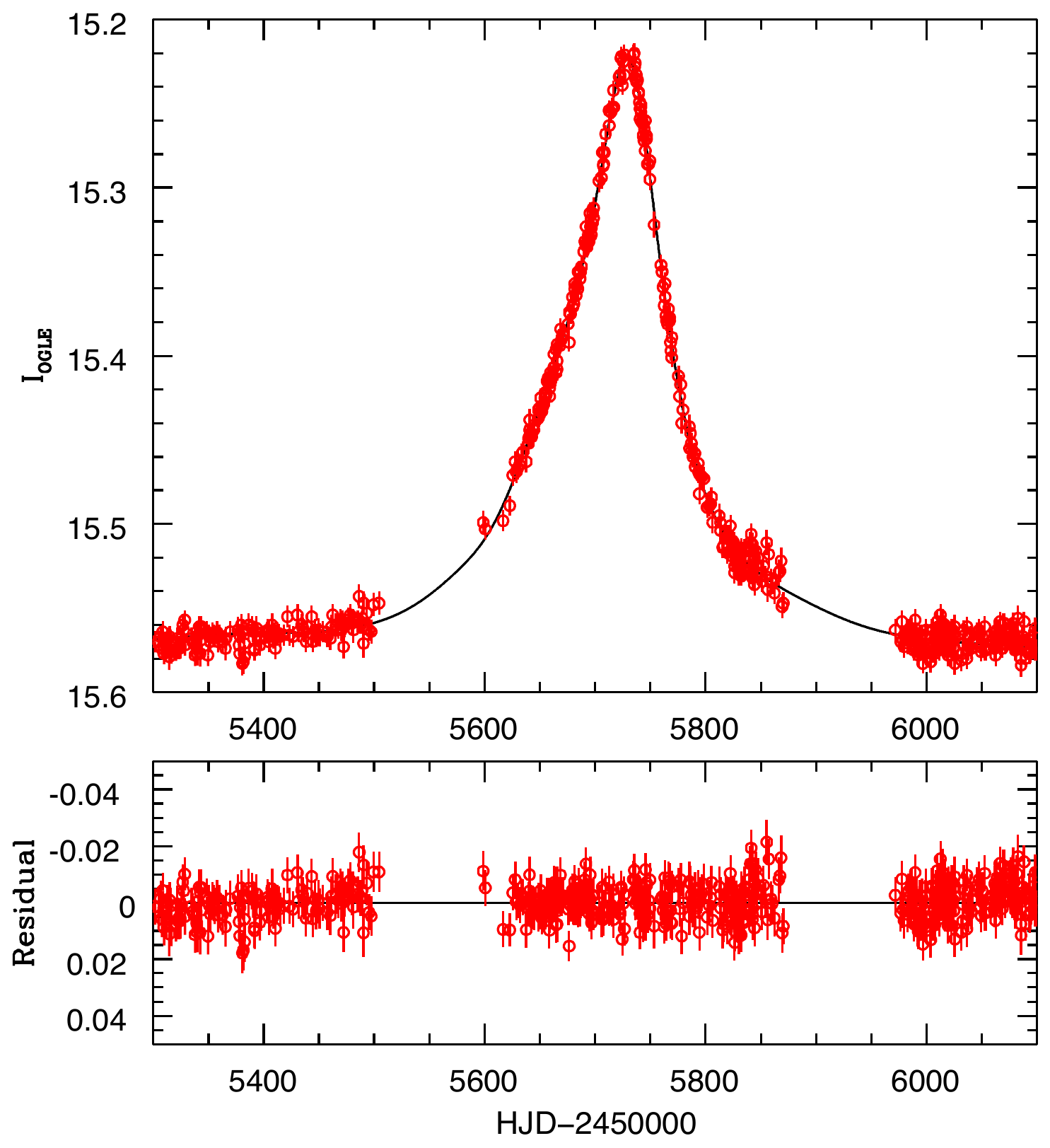}
\caption{\emph{Top:} Light curve of OB110125.  Red points indicate
  OGLE measurements.  The black line represents the best-fitting
  binary lens model, which includes microlensing parallax, but does
  not include binary orbital motion. \emph{Bottom}:  Residuals to
  best-fitting model.}
\label{fig:OB110125lc}
\end{figure}

Figure \ref{fig:OB110125lc} shows the OGLE $I$-band light curve as
well as the best-fitting photometric microlensing model for OB110125,
which has a static binary lens. The [wide, $\uominus$] solution is
preferred 
and shown in Figure \ref{fig:OB110125lc}.    Table
\ref{tb:parsOB110125} lists the posterior medians and 68\% confidence
intervals for each OB110125 model parameter.

Lens mass measurements of OB110125 are omitted because the first
astrometric observation took place too long after the event peak.
The light curve fitting favors a binary lens model with an Einstein
crossing time of $\sim$63 days, and the first observation of OB110125
was $\sim$ 347 days after the event peak (5.5 x $t_E$), at which point
the non-linear, lens-induced motion evolves too slowly to detect.   In
principle, the lens mass could still be estimated by a joint analysis
of the measured source proper motion, photometrically constrained
event parameters ($\tE$, $\piEvec$), and a Galactic model
\citep[e.g.]{Yee15}.  However, we choose to avoid relying on models of
the spatial and kinematic distributions of Galactic objects,
especially those of BHs, which carry large uncertainties.

%
%
\begin{deluxetable}{lc}[h]
\tabletypesize{\footnotesize}
\centering
\tablecaption{Microlensing Event Parameters - OB110125}
\tablecolumns{2}
\tablewidth{0.9\columnwidth}
\tablehead{
\colhead{Parameter} &
\colhead{Photometry}
}
\startdata
$t_0$ (HJD - 2450000)                         &      5724.43$^{+        0.72}_{-        0.70}$  \\ [0.2cm]
$u_0$                                         &       -0.611$^{+       0.031}_{-       0.030}$  \\ [0.2cm]
$t_E$ (days)                                  &         62.9$^{+         2.4}_{-         2.6}$  \\ [0.2cm]
$\pi_{E,E}$                                   &        0.515$^{+       0.034}_{-       0.033}$  \\ [0.2cm]
$\pi_{E,N}$                                   &       -0.437$^{+       0.036}_{-       0.037}$  \\ [0.2cm]
s                                             &         3.11$^{+        0.11}_{-        0.11}$  \\ [0.2cm]
q                                             &         1.07$^{+        0.16}_{-        0.17}$  \\ [0.2cm]
$\alpha$                                      &        3.517$^{+       0.008}_{-       0.009}$  \\ [0.2cm]
$I_{\mathrm{OGLE}}$                           &        15.54$^{+        0.16}_{-        0.17}$  \\ [0.2cm]
$f_{b}/f_{s}$                                 &       -0.035$^{+       0.129}_{-       0.160}$  \\ [0.2cm]
$\chi^{2}$                                    &       1012.7                                    \\ [0.2cm]
N$_{\mathrm{dof}}$                            &          995                                    \\ [0.2cm]
\enddata
\tablenotetext{}{}
\label{tb:parsOB110125}
\end{deluxetable}

\section{Discussion}

\label{sec:discussion}

Despite the unusually poor seeing conditions in which most of our data
were collected, and a time-sampling of lensing events that was far
less than optimal, we have demonstrated the feasibility of detecting
black holes via astrometric methods, given better observing conditions
and a well-timed set of observations over a longer (several year)
baseline.  While we have not yet detected a significant astrometric
microlensing signal, we achieved astrometric precisions in a few
epochs, when seeing was not poor, that would be sufficient to trace
the lens-induced astrometric shifts due to a black hole lens.

A binary lens model is favored for two of the three events, OB110022
and OB110125.  We measure the OB110022 combined
lens mass to better than $\sim$50\%.   This mass constraint is likely
comparable to those achieved by using finite source effects to
estimate the source angular size, $\thetastar$ \citep[e.g.][]{Park13,
  Park15, Jung15, Zhu15}. 
One event, OB120169, is consistent with an isolated lens, which
remains a BH candidate.  The best-fitting astrometric models for both
$\uoplus$ and $\uominus$ solutions of OB120169 predict that the
y-component of the source apparent motion may be turning over and
approaching that of the true (unlensed) motion.  Follow-up NIRC2
observations in 2015-2016 will be a valuable test of this model and
significantly constrain the lens mass posteriors.

To date, the most precise mass measurement of an isolated unseen lens without
finite source effects is $\sim$30\% for event OGLE-2014-BLG-0939
\citep[$\tE \approx$ 24 days][]{Yee15}.   This was part of a pilot
100-hour Spitzer campaign to precisely measure microlensing parallaxes
of 21 events via simultaneous Spitzer and ground-based telescopes
observations \citep{Udalski15, CalchiNovati15}.  While better
microlensing parallaxes are a key ingredient to improving lens mass
constraints, isolated lens mass measurements are still limited by
their dependence on Galactic models and unknown $\murelvec$; and the
uncertainties are difficult to quantify.

The lens mass constraints presented here, from a combination of
photometry and astrometry, do not depend on Galactic models.  This
completely ground-based technique is promising, but the observations
and astrometric modeling techniques used in this pilot attempt can be
refined to achieve better lens mass constraints.  In future, it would
be useful to construct an analysis that incorporates the available
photometry and astrometry simultaneously.  Moreover, future
astrometric models should include the effects of parallax and
binarity.  Once microlensing event parameters have been constrained,
Galactic models might serve as a sanity check.

\section{Future Experimental Design}
\label{sec:future}

This study constitutes the first lens mass constraint from
ground-based astrometry.  It served as a pilot attempt, intended to
test the feasibility of detecting astrometric microlensing, and
suffered from unusually poor observing conditions during the critical
epoch around $t_0$.  We now explore the optimization of future
searches for isolated BHs assuming astrometric precision typical for
average NIRC2 observing conditions, and similar photometric
constraints.

The ideal time sampling of astrometric observations is that which best
captures the non-linearity of the lens-induced astrometric shift,
distinguishing a lensing event from ordinary proper motion.  We first
discuss the qualitative characteristics of such an observing strategy,
considering practical limitations.  We then construct simulations of
various feasible observing strategies to quantitatively determine that
which would most significantly and efficiently constrain the lens
mass.

Assuming regular monitoring of light curves, the Einstein crossing
time of a lensing event, $t_{E}$ can be tightly constrained as the
photometric peak approaches.   If $t_{E} \gtrsim$ 50 days, and the
magnification increases sharply, characteristic of a low impact
parameter, one has a strong case for commencing astrometric
observations as soon as possible.  This first observing season is most
crucial because the lens is at closest approach to the source --- the
lensing signature is evolving most rapidly, and deviates the most from
unlensed (i.e. linear) motion.  Relatively dense sampling is required
to detect this evolution with high significance.  If observations
begin early enough before the photometric peak the $\sim180\degree$
change in direction of the shift can be captured. The timescale of
some events might be long enough ($t_E \gtrsim$ 200) to warrant
additional heavy observations at the beginning of year 2.  Follow-up
observations in subsequent years are not nearly as time sensitive
because the source apparent motion is approximately linear as the
lens-induced component decays asymptotically to zero.

With these ideas in mind, we evaluate a suite of observing strategies
via computer simulation.  To assess each strategy, we apply our
event-fitting methods to obtain lens mass posteriors from a synthetic
astrometric dataset.  The data are constructed from an astrometric
model of a typical lensing event sampled according to each particular
strategy, with injected noise comparable to our typical astrometric
precision (0.15 mas), assumed to be Gaussian.  We adopt photometric
priors representative of those obtained for OB110022, OB110125 and
OB120169.  The set of tested observing strategies is parameterized by:

\begin{enumerate}
\begin{item}
Number of observations in year 1 ($N_1$)
\end{item}
\begin{item}
Number of observations in each subsequent year ($N_{>1}$)
\end{item}
\begin{item}
Total number of consecutive years in which target is observed ($N_{\mathrm{yrs}}$)
\end{item}
\end{enumerate}

\begin{figure*}
\centering
\includegraphics[width=14cm]{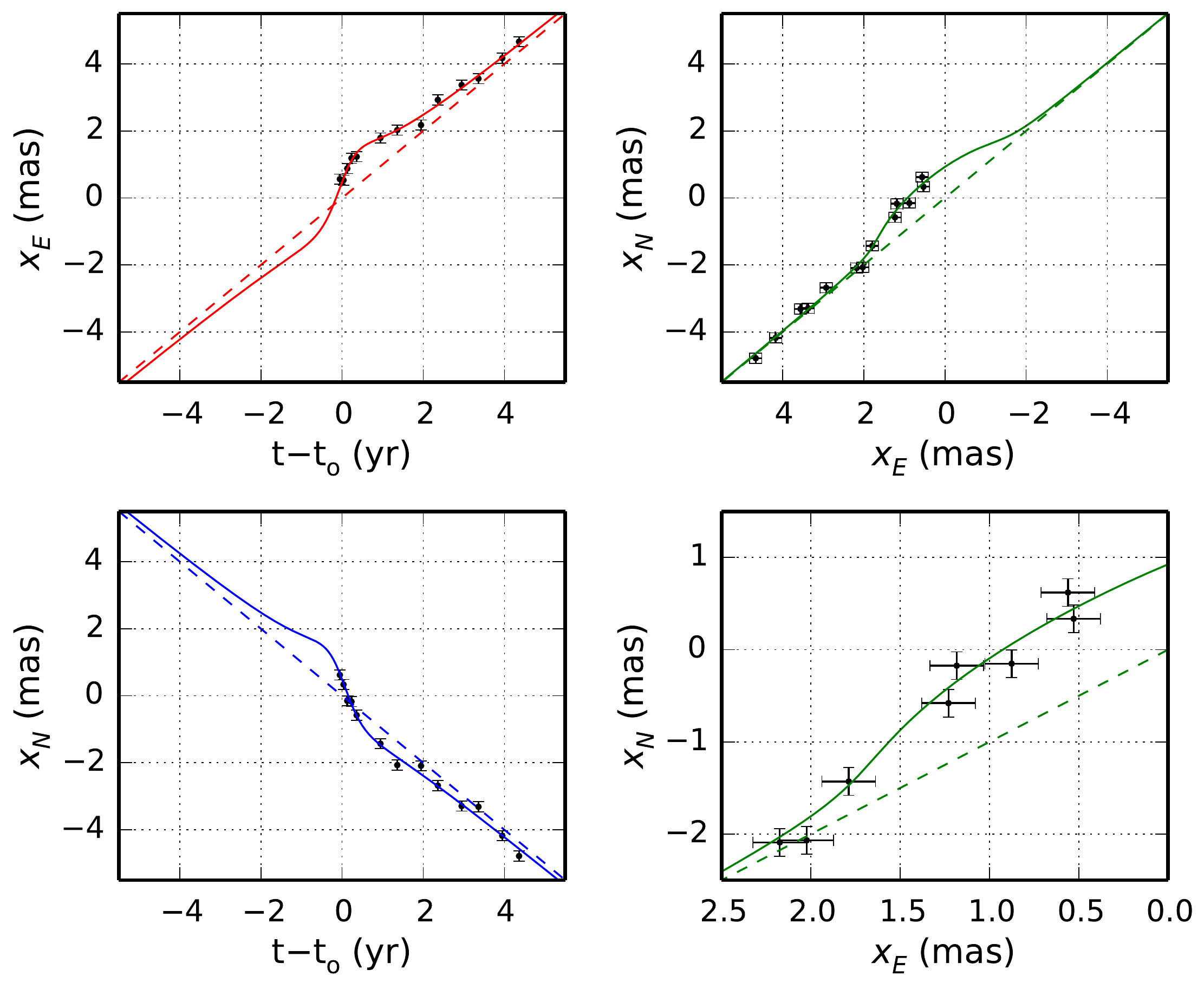}
\caption{A simulation of the 2D astrometric shift due a 10-M$\subsun$
  black hole at 4 kpc microlensing a background source at 8 kpc with a
  relative proper motion of 7 mas yr$^{-1}$ and impact parameter $\uo$
  = 0.5. The model X (\emph{top left}) and Y (\emph{bottom left})
  positions vs. time are shown (\emph{solid}) overlaid with simulated
  astrometric measurements and an unlensed source motion
  (\emph{dashed}). The measurements on the plane of the sky are shown
  in the top right and zoomed in on the bottom right. Error bars
  represent the 1$\sigma$ astrometric uncertainty expected from NIRC2
  observations (0.15 mas).  The astrometric effects of parallax have
  been omitted.}
\label{fig:ObsStrat}
\end{figure*}

\begin{figure}
\centering
\includegraphics[width=9cm]{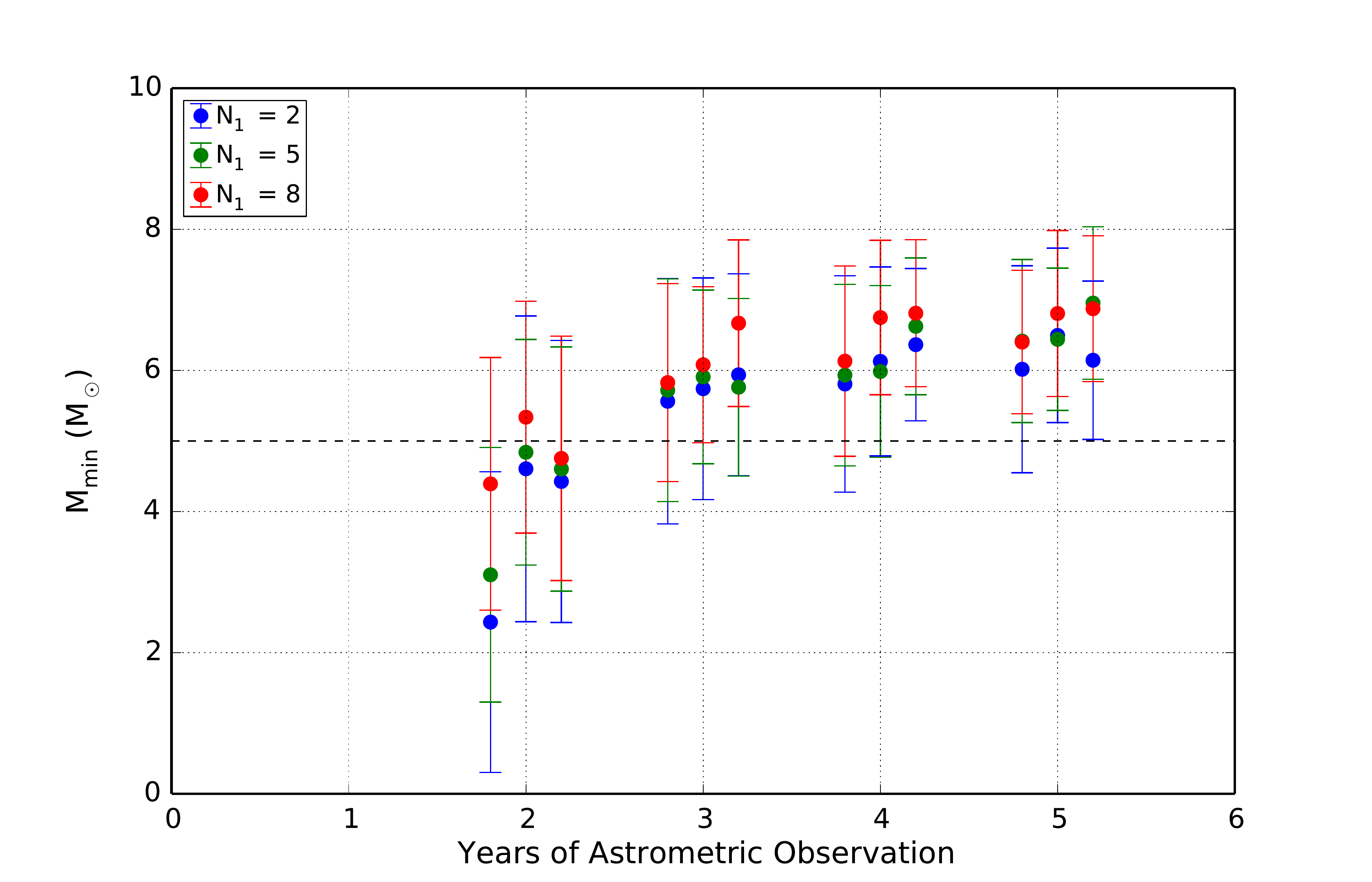}
\caption{3$\sigma$ lower limits for lens mass ($M_{\mathrm{min}}$)
  recovered from different simulated observing strategies assuming a
  10-M$\subsun$ lens at 4 kpc, and a source at 8 kpc. Colors
  distinguish the number of nights of observations in year 1 ($N_{1}$)
  as labeled in the legend. The horizontal axis indicates the total
  number of years of observation ($N_{\mathrm{yrs}}$).  For each value
  of $N_{\mathrm{yrs}} >1 $ there are three columns, corresponding to
  (left-to-right) 1, 2, and 3 observations per year after year 1
  ($N_{>1}$).  The $M_{\mathrm{min}}$ values plotted for each
  observing strategy indicate the median of 100 trials, with
  1-$\sigma$ error bars.  The dashed line indicates 5 M$\subsun$---the
  minimum mass that we assume is required to confidently conclude a
  BH.}
\label{fig:MassSens}
\end{figure}

We test all permutations of the allowed parameter values,
$N_1=\left\{2, 5, 8\right\}$, $N_{>1}=\left\{1, 2, 3\right\}$ and
$N_{\mathrm{yrs}}=\left\{2, 3, 4, 5\right\}$, comprising 36 different
observing strategies.  We assume the target is located toward the
Galactic Bulge, observable from the ground between April 1 and August
31 of each year. For simplicity, the dates of first and last
observations in each year are always April 1 and August 31
respectively, and the remainder are spaced in time such that the
astrometric signal is evenly sampled.  The time of minimum separation
is assumed to be 20 days after the first observation.  We assume a
10-M$\subsun$ lens at 4 kpc, a source at 8 kpc, and relative
source-lens motion of 4 mas yr$^{-1}$.  The astrometric model for this
event is shown in Figure \ref{fig:ObsStrat}, along with a set of
simulated measurements corresponding to a the strategy ($N_1$,
$N_{>1}$, $N_{\mathrm{yrs}}$) = (5, 2, 5).   We identify the 3$\sigma$
lower limit of the lens mass, $M_{\mathrm{min}}$ from the
corresponding marginalized posterior.  One can safely conclude a black
hole lens if $M_{\mathrm{min}}~ \gtrsim 5$ M$\subsun$.   We simulate
each observing strategy 100 times to obtain a distribution of results
representative of 0.15 mas astrometric noise and plot the median
$M_{\mathrm{min}}$ values and 1$\sigma$ uncertainties in Figure
\ref{fig:MassSens}.

Within this margin of error, none of the two-year observing strategies
confirm $M > 5$ M$\subsun$ with 3$\sigma$ confidence.  A minimum
$N_{\mathrm{yrs}} = 3$ is required.  There is significant improvement
when $N_{\mathrm{yrs}}$ increases from two to three, but further
increases have minimal and diminishing returns.  Figure
\ref{fig:ObsStrat} shows that astrometric measurements both two and
three years after the event peak adequately constrain the unlensed
proper motion.  Strategies with $N_{>1} \geq 2$ are recommended to
protect against single measurement outliers.   Adopting $N_1 >  2$
marginally improves mass constraints, but it does not seem to be
necessary.  Still, we recommend  $N_1 \geq  3$ to protect against
single measurement outliers.   In summary, we conclude that future
searches for isolated BHs should adopt an observing strategy with
$N_1 \geq  3$,  $N_{>1} \geq 2$, and $N_{\mathrm{yrs}} \geq 3$.

Finally, our use of photometry to identify candidate long-duration
events for astrometric follow-up might favor binary lenses, which must
be considered in future target selection.  In principle, a goodness of
fit comparison of the single-lens and multiple-lens models before
astrometric follow-up might be a viable screening option.
Unfortunately, the primary factor distinguishing these models is the
degree of symmetry in the light curve around the peak.  Therefore, it
is unlikely that the lens multiplicity can be constrained before
astrometric follow-up begins (i.e. before or near the peak).
Exceptions include caustic crossing events for which finite-source
effects appear before the event peak.

\section{Conclusions}
\label{sec:conclusions}

While stellar mass black holes still remain elusive, this study
demonstrates the feasibility of detecting them in the near future, and
provides a strategy to do so efficiently based on demonstrated
astrometric and photometric precision.  This is the first study that
uses both astrometric and photometric ground-based measurements to
constrain lensing event parameters.  The value of a precise
measurement of the proper motion of the source or the combined
source-lens system is demonstrated in the tight upper limit on the
mass of OB110022, which we show is not a black hole. Moreover, by
uncovering and mitigating Keck/NIRC2 systematic effects like spatial
variation of the PSF, and exploring the effects of using different
transformations in cross-epoch alignment, this study informs future
analysis of any astrometric data, regardless of the scientific
motivation.  Follow-up astrometric observations of one of our targets,
OB120169, should further constrain mass of the lens, which could be a
black hole. Very few studies have attempted the astrometric detection
of stellar mass black holes, but it seems a worthwhile endeavor.

\section{Acknowledgements}
The authors would like to thank Subo Dong (Kavli Institute for 
Astronomy and Astrophysics, Peking University) for his significant
contributions to this work. 
J.R.L. acknowledges support for this work from the California
Institute of Technology Millikan Postdoctoral Fellow program and the
NSF Astronomy and Astrophysics Postdoctoral Fellow program
(AST−1102791).
E.O.O. is an incumbent of the Arye Dissentshik career development
chair and is grateful for support by grants from the  Willner Family
Leadership Institute Ilan Gluzman (Secaucus NJ), Israel Science
Foundation, Minerva, Weizmann-UK, and the I-Core program by the
Israeli Committee for Planning and Budgeting and the Israel Science
Foundation (ISF).
The OGLE project has received funding from the National Science
Centre, Pland, grant MAESTRO 2014/14/A/ST9/00121 to AU.
E.S. acknowledges the SWOOP writing retreat and its participants for useful feedback.
The data presented herein were obtained at the W. M. Keck Observatory,
which is operated as a scientific partnership among the California
Institute of Technology, the University of California and the National
Aeronautics and Space Administration. The Observatory was made
possible by the generous financial support of the W. M. Keck
Foundation. The authors wish to recognize and acknowledge the very
significant cultural role and reverence that the summit of Mauna Kea
has always had within the indigenous Hawaiian community. We are most
fortunate to have the opportunity to conduct observations from this
mountain.

\clearpage
\newpage
\bibliographystyle{apj}
\bibliography{ms.bib}

\appendix

\section{A. Weighting Schemes in Cross-Epoch Alignment}
\label{sec:weighting}
We experimented with four different weighting schemes to apply to the
sources used to derive the cross-epoch transformation.   We tried
applying weights for each star in each epoch, $i$, given by,
\begin{enumerate}
\item $W_{i} = 1$
\item $W_{i} = 1 / \sigma_i^2$
\item $W_{i} = 1 / \sqrt{\sigma_i^2 + \sigma_0^2 + (\sigma_\mu[t-t_0])^2}$
\item $W_{i} = 1 / \left[\sigma_i^2 + \sigma_0^2 + (\sigma_\mu[t-t_0])^2\right]$
\end{enumerate}
where $\sigma_i$ is a star's positional uncertainty in epoch $i$,
$\sigma_0$ is the positional uncertainty in the reference epoch (April
2013), $\sigma_\mu$ is its proper motion uncertainty and $t-t_{0}$ is
the time between the reference epoch and epoch of interest.  Including
$\sigma_0$ and $\sigma_\mu (t-t_0)$ incorporates the fact that, since
all coordinate systems are aligned to the reference epoch, the
positional uncertainty grows with time from that epoch in proportion
to the source velocity.   We applied these weights averaged over $x$
and $y$. These four weighting schemes are hereafter denoted W1, W2, W3
and W4 respectively.  W3 and W4 incorporate positional uncertainties
associated with time from the reference epoch.  W3 results in a
weighting that is less sensitive to differences in uncertainty than
W4.

\begin{figure}
\centering
\subfigure[W1]{
\includegraphics[width=6in]{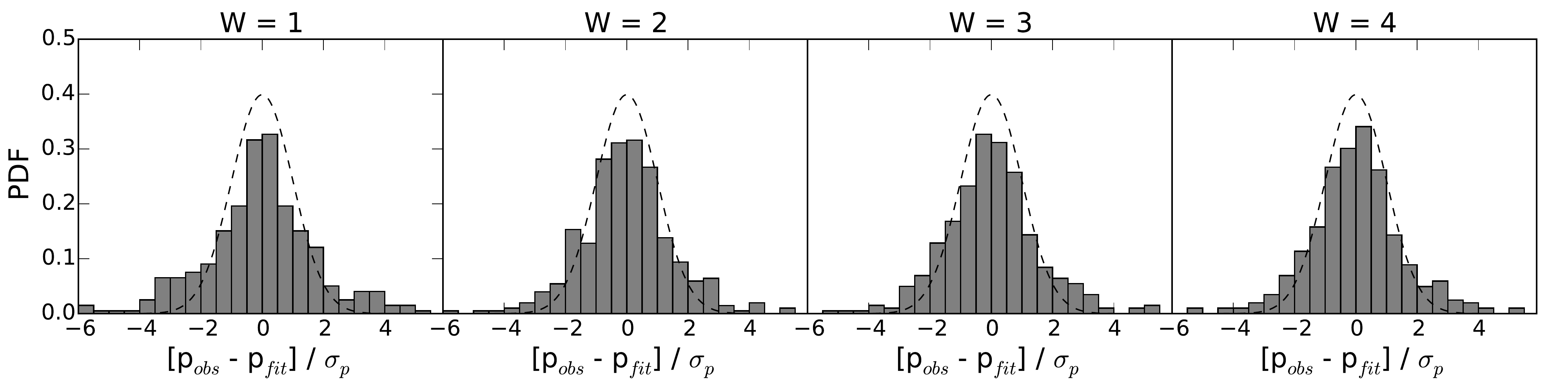}}
\caption{Histogram of residuals to the proper motion fits for all
  stars in the OB110022 epochs, using the four different cross-epoch
  alignment weighting schemes. The residuals are normalized by the 
  positional uncertainties. Their mutual Gaussian shape indicates
  that the choice of weighting scheme does not affect the errors on
  the transformations and proper motion fits.}
\label{fig:wresall}
\end{figure}

\begin{figure}
\centering
\includegraphics[width=8cm]{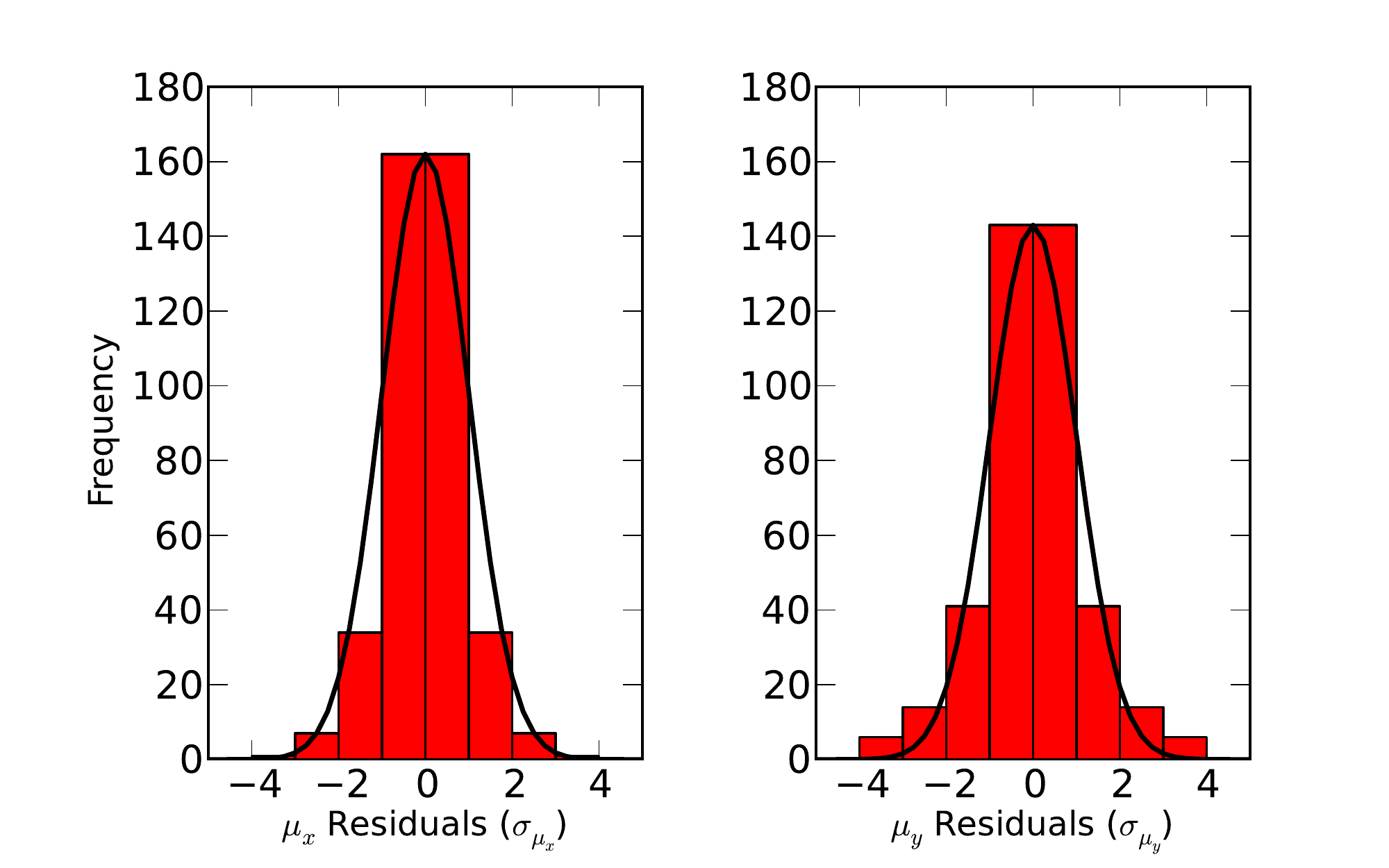}
\caption{Distribution of the difference between proper motion
  measurements of each star in the OB110022 field across different
  cross-epoch alignment
  weighting schemes, in units of their associated error.  The Gaussian
  distribution indicates that the derived proper motions are
  independent of weighting scheme.}
\label{fig:VsigDist}
\end{figure}

Figure \ref{fig:wresall} shows the distribution of position residuals obtained
using the four different weighting schemes for OB110022, normalized
by the positional uncertainy, $\sigma_{pos}$.   Their
consistency suggests that our particular choice of weighting scheme
does not affect the precision of the proper motion fit.  However, we
must still ensure that its accuracy is also invariant of the weighting
scheme.  Even if the choice of weighting scheme does not influence the
error of the proper motion fit it could still affect the best fit
value.  To test this we obtain the difference distribution $\delta$ of
the proper motion measurements for each source $q$ under weighting
scheme $p$ versus  a different weighting scheme $k$ and check that it
is consistent with the errors on each proper motion measurement.
Specifically,

\begin{equation}
\delta_{q,p,k} = \frac{\mu_{q, p}-\mu_{q, k}}{\left(\sigma_{\mu_{q,
        p}}^{2} + \sigma_{\mu_{q, k}}^{2}\right)^{1/2}},
\end{equation}
for all stars $q$, where $p \ne k$.  Figure \ref{fig:VsigDist} shows
the resulting distributions obtained in both x and y.  Given the
apparent consistencies of both this proper motion difference
distribution and the position residuals distributions with their
associated errors, we conclude that the our particular choice of
weighting scheme will not affect our results.  Therefore, we
arbitrarily elected to use W4, which is the statistically appropriate
weighting scheme when errors are normally distributed and well
characterized.

%
%
%

\section{B. Transformation Order in Cross-Epoch Alignment}
\label{sec:align_order}
We tested cross-epoch transformations of order O$=$1, 2, and
3 in a manner similar to \citet{Clarkson:2012}. 
For each order, we define the metric $\chi_{aln,vel}^2$(O), as a means of quantifing the 
goodness-of-fit for the combined transformation fits and proper motion 
fits for all stars over all epochs. We sum the residuals
between the observed position, after transformation, and the predicted position from the 
best-fit proper motion for all epochs and all stars in each target's field:
\begin{equation}
\chi_{aln,vel}^2 = \sum_s^{N_{\mathrm{stars}}} \sum_e^{N_{\mathrm{epochs}}} \left[\frac{x_{\mathrm{obs,s,e}} -
  x_{\mathrm{fit,s,e}}}{\sigma_{\mathrm{x,s,e}}}\right]^2
+ 
\left[\frac{y_{\mathrm{obs,s,e}} - y_{\mathrm{fit,s,e}}}{\sigma_{\mathrm{y,s,e}}}\right]^2
\end{equation}
where $\sigma_{\mathrm{x,s,e}}$ and $\sigma_{\mathrm{y,s,e}}$ include only 
the positional errors, $\sigma_\mathrm{p}$, and do not include the alignment 
errors.
The number of degrees of freedom (DOF) for the combined alignment and velocity fits
is the difference between $N_{\mathrm{data}}$, the total number of positional measurements,
and $N_{\mathrm{par}}$, the number of free parameters including
those from the transformation for each epoch and the proper motion
fits for all stars in the field of view (Table \ref{tab:align_ftest}). 
For increasing model complexity (i.e. polynomial order), 
we evaluated the change in $\chi_{aln,vel}^2$ using the F-ratio,
\begin{equation}
F = \left(\frac{\chi^2(O-1) - \chi^2(O)}{\chi^2(O)}\right) 
\frac{N_{\mathrm{data}} - N_{\mathrm{par}}(O)}{N_{\mathrm{par}}(O) -  N_{\mathrm{par}}(O-1)}.
\end{equation}
Table \ref{tab:align_ftest} gives the resulting $\chi_{aln,vel}^2$
values for each order, the F-ratios 
when increasing from $O-1 \rightarrow O$,
and the p-value, which is the
probability of obtaining this F-ratio, or higher, from chance. 
Lower p-values indicate more significant benefits to increasing 
the order of the transformation polynomials.
We note that the $\chi_{aln,vel}^2$ values are high relative to the degrees
of freedom, which typically suggests that uncertainties should be
re-scaled to larger values. However, the F-ratio is insensitive 
to error re-scaling and can still be used to select the optimal order
of the transformation. 
For OB110022, the $\chi_{aln,vel}^2$ values in Table \ref{tab:align_ftest} increase 
for higher-order fits, which may be caused either by instability 
of the high-order fit, given the small number of epochs, or a high condition number
in the inversion. 
Furthermore, $\chi_{aln,vel}^2$ values do not
include the transformation uncertainties and should not be interpreted
as the final quality of our astrometric transformations and proper
motion fits. 
OB110022 showed
no significant improvement when advancing to 2nd order (F-ratio$<$1),
therefore we adopt O$=$1 for this source. 
Both OB110125 and OB120169 showed marginal improvement going from 
O$=1\rightarrow2$ and
the $\chi^2$ distribution for O$=$2 showed significant improvement,
therefore we adopted O$=$2 for these sources.

\begin{figure}
\centering
\includegraphics[width=4in]{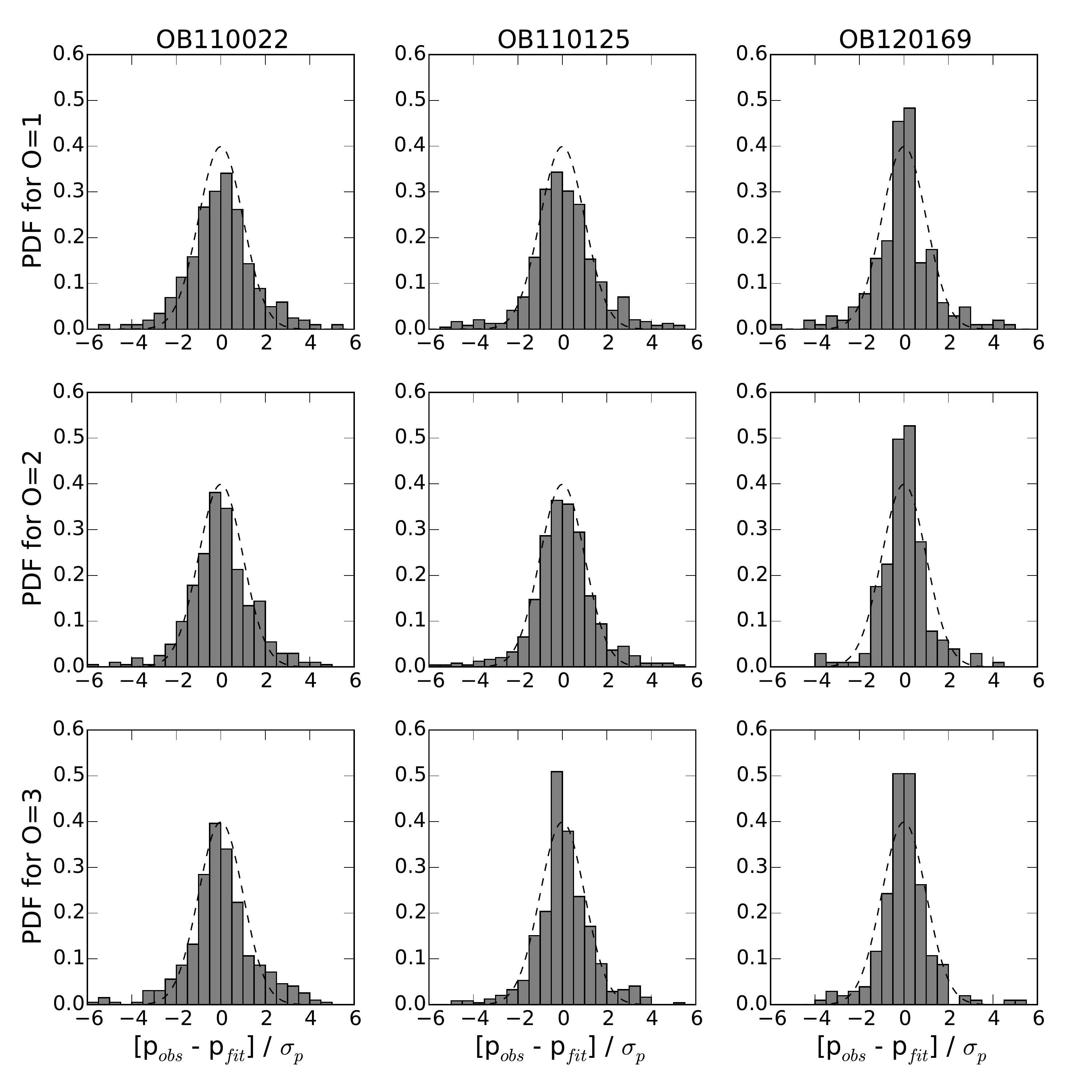}
\caption{
Histogram of error-weighted residuals to the linear source proper
motion fits, $(p_{obs} - p_{fit}) / \sigma$,
for all three targets and different transformation orders, O=1, 2, 3. 
The errors include only positional errors, $\sigma_p$. Residuals 
for both X and Y are included. 
The resulting distribution of residuals are largely consistent with
with a normal distribution with a 1$\sigma$ spread for all orders 
({\em black dashed line}).
Note the final analysis uses O=1 for OB110022 
and O=2 for OB110125 and OB120169.
}
\label{fig:hist_resid_p}
\end{figure}

\begin{figure}
\centering
\includegraphics[width=4in]{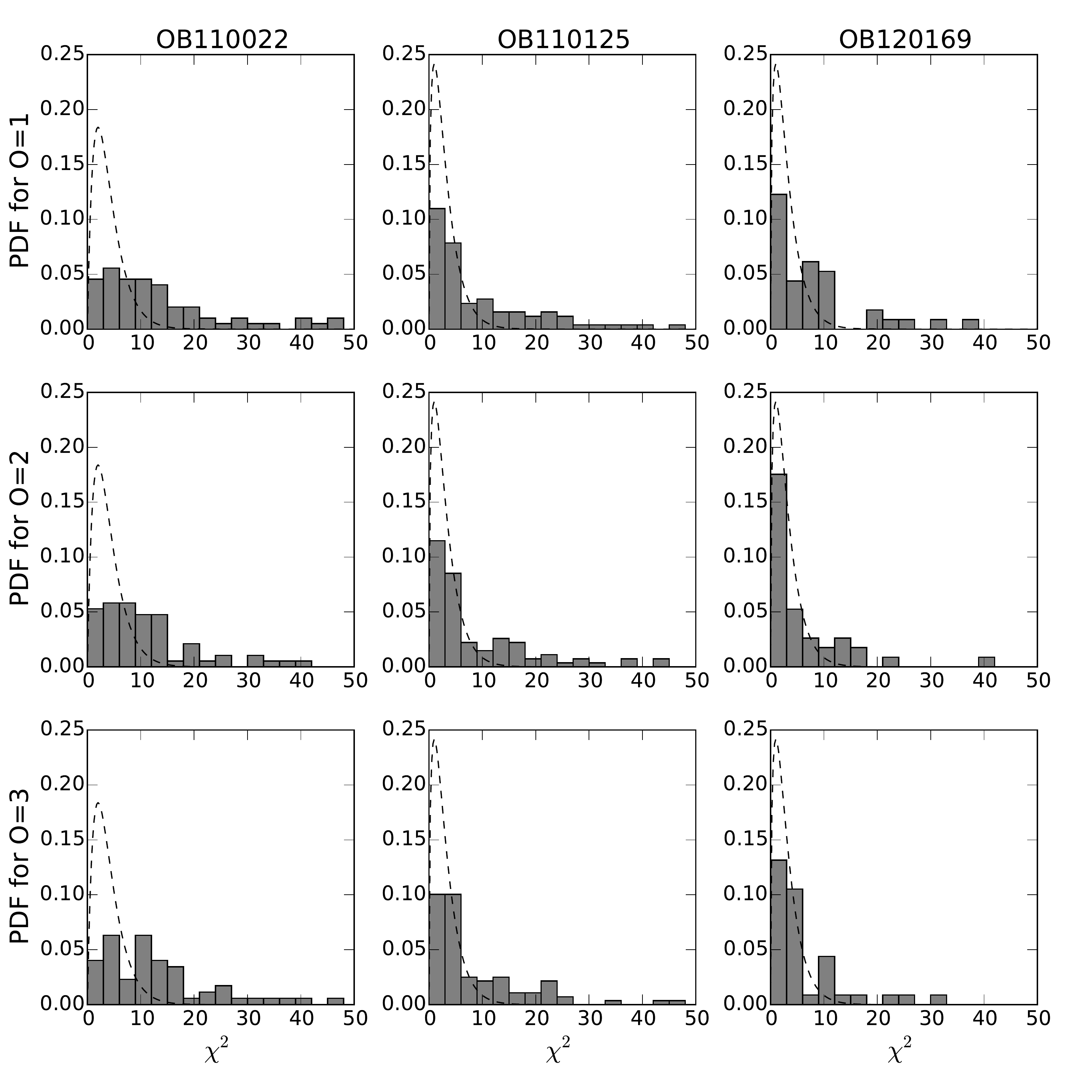}
\caption{
Histogram of $\chi^2$ values of each star's proper motion fit
for all three targets and different transformation orders,
O=1, 2, 3. The $\chi^2$ values for both X and Y are included
in the histogram and the degrees of freedom for the expected
$\chi^2$ distribution ({\em black dashed line}) is given by the number
of epochs of data - 2 free parameters in the velocity fit. 
Only positional errors, $\sigma_p$, were included.
The F-test used to evaluate the transformation order sums these 
$\chi^2$ values over the number of stars and X and Y and the
final degrees of freedom is modified to include the transformation
parameters for each epoch of data. 
OB110125 and OB120169 show statistically significant improvements 
in the $\chi^2$ distribution relative to the expectation 
when going from O$=1\rightarrow 2$.
}
\label{fig:hist_chi2_p}
\end{figure}

\begin{figure}
\centering
\includegraphics[width=4in]{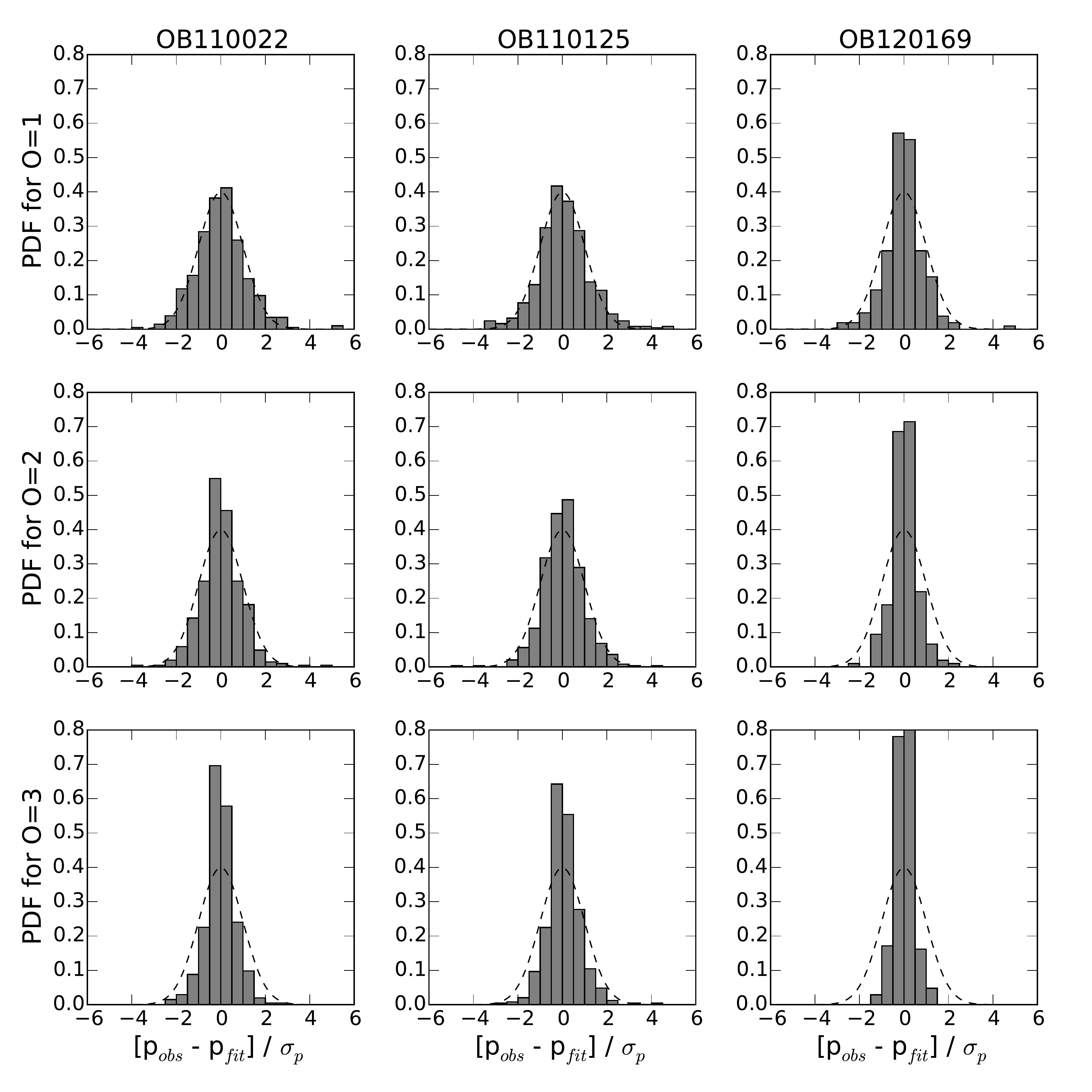}
\caption{
Identical to Figure \ref{fig:hist_resid_p} only the 
errors include both positional and alignment errors, $\sigma =
\sqrt{\sigma_p^2 + \sigma_a^2}$.
The resulting width of the residuals decreases with increasing
transformation order as the alignment errors increases.
}
\label{fig:hist_resid_pa}
\end{figure}

\begin{figure}
\centering
\includegraphics[width=4in]{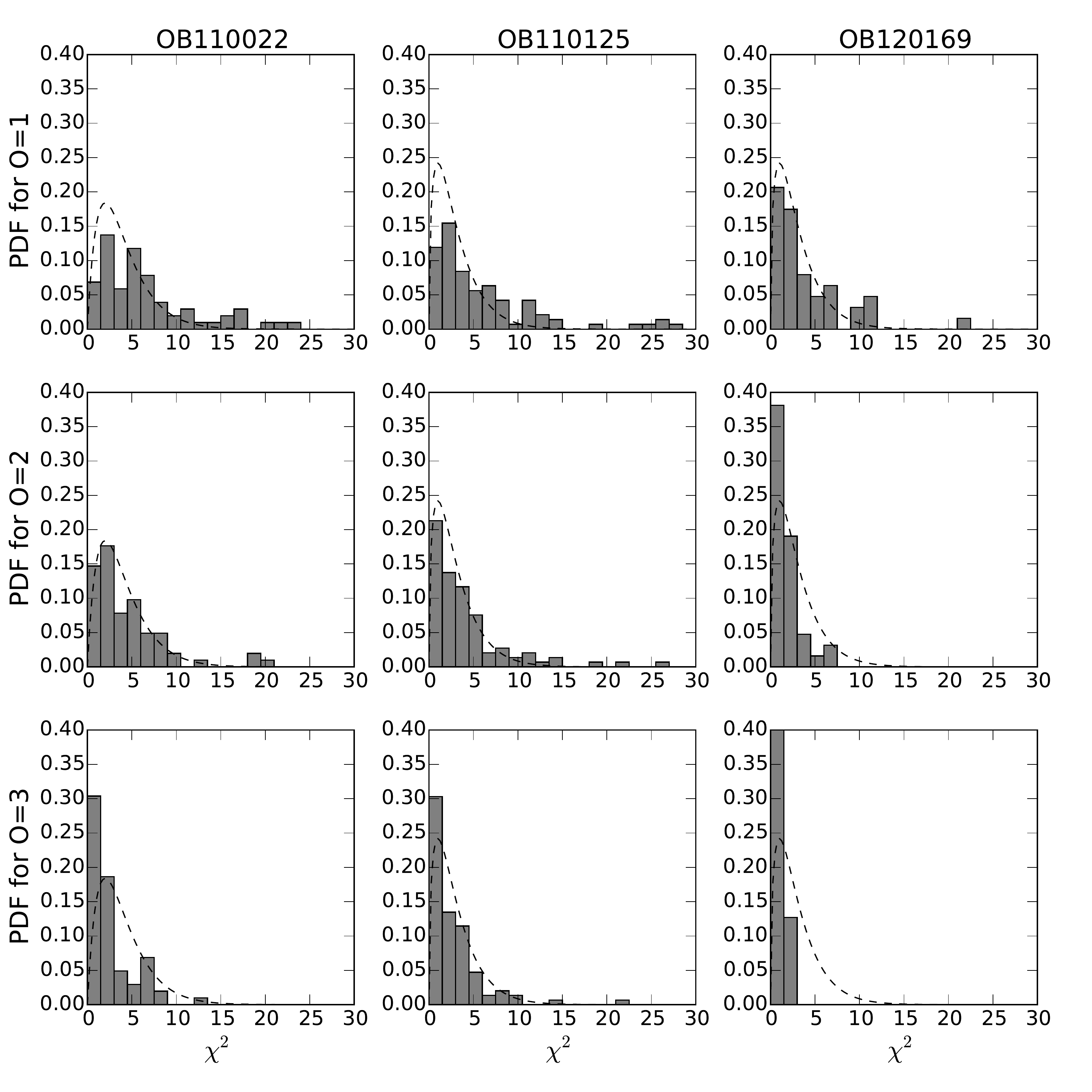}
\caption{
Identical to Figure \ref{fig:hist_chi2_p} only the 
errors include both positional and alignment errors, $\sigma =
\sqrt{\sigma_p^2 + \sigma_a^2}$.
The resulting $\chi^2$ distribution shows only slight
changes for OB110022 with increasing transformation
order. OB110125 and OB120169 show more dramatic improvements when changing from 
O$=1 \rightarrow 2$.
}
\label{fig:hist_chi2_pa}
\end{figure}

Figure \ref{fig:hist_resid_p} shows the distribution of 
residuals for all stars in all epochs for each target field and each 
transformation order and Figure \ref{fig:hist_chi2_p} shows the distribution
of $\chi_{vel}^2$ for each stars proper motion fit 
(i.e. weighted residuals summed over $N_{epochs}$). Note that these
$\chi_{vel}^2$ values do not include the alignment error; but, are the
individual values (one for each star) that go into the F-ratio.
Again, the relative improvement is most significant for 
O$=1 \rightarrow 2$ for OB120169 and OB110125. Note that these distributions of 
residuals and $\chi_{vel}^2$ values are only used to judge the choice of 
transformation order and are not final as they do not include
the uncertainties in the transformation process. 
For completeness, the distribution of residuals and $\chi_{vel}^2$ values
including both the positional and alignment errors are shown in 
Figures \ref{fig:hist_resid_pa} and \ref{fig:hist_chi2_pa} for 
all three targets and all three transformation orders. 

Finally, the microlensing fits were run for O$=$1 and O$=$2 for
OB110022 and OB120169 and there
was a negligable change in the final lens-mass posteriors of less than
$<$0.1 \msun for all limits.

%
%
\begin{deluxetable}{llrrrrlrr}
\tabletypesize{\footnotesize}
\tablecaption{F-ratio for Alignment Order}
\tablecolumns{9}
\tablewidth{0pt}
\tablehead{
 & & & & & & Order & & \\
Source & Order & $\chi_{aln,vel}^2$\tablenotemark{a} & N$_{data}$ & N$_{par}$ & DOF$_{aln,vel}$ & 
Change & F-ratio & p-value\tablenotemark{b}
}
\startdata
OB110022 & O$=1$ & 1023 & 408 & 166 & 242 & & & \\
         & O$=2$ & 1156 & 408 & 196 & 212 & O$=1 \rightarrow 2$ & -0.81 & 1.0000 \\
         & O$=3$ & 2487 & 408 & 236 & 172 & O$=2 \rightarrow 3$ & -2.30 & 1.0000 \\
OB110125 & O$=1$ & 2784 & 500 & 224 & 276 & & & \\
         & O$=2$ & 1995 & 500 & 248 & 252 & O$=1 \rightarrow 2$ &  4.15 & 0.0000 \\
         & O$=3$ & 1810 & 500 & 280 & 220 & O$=2 \rightarrow 3$ &  0.70 & 0.8822 \\
OB120169 & O$=1$ & 1277 & 210 & 108 & 102 & & & \\
         & O$=2$ & 926  & 210 & 132 &  78 & O$=1 \rightarrow 2$ &  1.23 & 0.2403 \\
         & O$=3$ & 787  & 210 & 164 &  46 & O$=2 \rightarrow 3$ &  0.25 & 0.9999 \\
\enddata
\tablenotetext{a}{The value reported here only includes
  positional uncertainties and not transformation errors. It is only
  used to judge the relative improvement in the fit with changes to
  the transformation order and should not be taken as an indication of
  the final quality of the proper motions.}
\tablenotetext{b}{Lower p-values indicate more significant benefits
to increasing the order of the transformation polynomials.}
\label{tab:align_ftest}
\end{deluxetable}

%
%
%

%
%
%

%
%
%

%
%
%

\end{document}